\documentclass[12pt]{article}

\usepackage{fullpage,graphics,color,rotating}

\newcommand{\D}{\displaystyle}

\definecolor{dark}{gray}{.8}
\definecolor{light}{gray}{.9}
\begin{document}

\begin{titlepage}
\setcounter{page}{0}
\thispagestyle{empty}
\renewcommand{\thefootnote}{\fnsymbol{footnote}}

\begin{flushright}
ICN-UNAM-98-06
\end{flushright}
\vspace{0.3in}

\begin{center}
\LARGE\bf Glueball Spectrum from an Effective Hamiltonian \\[0.25in]
\large\rm P.O. Hess\footnote{On sabbatical leave from the Instituto de Ciencias
Nucleares, UNAM, Mexico; \protect\\
e-mail address: hess@theo.physik.uni-giessen.de} \\[2.5mm]
\normalsize\it Institut f\"ur Theoretische Physik,
Justus-Liebig-Universit\"at, \\[0.5mm]
Heinrich-Buff-Ring 16, D-35392 Giessen, Germany \\[4.5mm]
\large\rm A. Weber, C.R. Stephens, S.A. Lerma H.,
J.C. L\'opez V.\footnote{e-mail addresses: axel@nuclecu.unam.mx,
stephens@nuclecu.unam.mx, alerma@nuclecu.unam.mx, \protect\\
vieyra@pythia.nuclecu.unam.mx} \\[2.5mm]
\normalsize\it Instituto de Ciencias Nucleares, UNAM, \\[0.5mm]
Circuito Exterior C.U., A.P. 70-543,
Delegaci\'on Coyoac\'an, \\[0.5mm]
04510 M\'exico D.F., Mexico \\[4.5mm]
\large\rm December 22, 1998
%\today
\end{center}
\vspace{0.25in}

\begin{abstract}
Using simple and general arguments we propose an effective Hamiltonian
for the description of low-energy pure QCD. The Hamiltonian is a function
of spatially constant collective modes. Its eigenstates can be organized into
bands classified by the irreducible representations of an $O(8)$ group. The
latter also determine parity and charge conjugation of the states. The energy
spectrum agrees well with the glueball spectrum as measured on the lattice,
and in particular the level ordering with respect to spin is naturally
explained.
\end{abstract}

\end{titlepage}
\setcounter{footnote}{0}

\section{Introduction}

QCD is widely considered to be the fundamental theory governing the
strong interactions. Its chief success to date has been in describing
experimental data at high energies --- deep inelastic scattering ---
where the partonic nature of hadronic constituents plays a dominant r\^ole.
In this regime the asymptotically free nature of the QCD running coupling
constant, $g$, comes to the fore allowing for a perturbative expansion
using as basis states free, i.e. $g\rightarrow0$, gluons and quarks.
At low energies, however, comparable succcess has been conspicuous by its
absence. As is well known this is due to the phenomenon of infrared
slavery which has as a consequence the confinement of coloured states, i.e.
gluons and quarks are bound together to form hadrons.
How to describe the crossover as a function of scale between such
radically different effective degrees of freedom remains as one of the
most challenging problems in particle physics.

Without reliable access to the infrared via the fundamental theory
itself recourse is often made to treatments that use additional assumptions,
such as adding confinement by hand (as in Refs.\
\cite{mitbag}--\cite{swanson}), or that
use effective field theory descriptions such as working
directly with the meson and baryon fields and adjusting the
interaction parameters
to experimental data \cite{efftheo}. The fundamental question remains however:
how does one derive the effective theory consistently from the underlying
more fundamental theory? In the present case, as QCD has only one parameter,
$\Lambda_{\scriptstyle QCD}$, how do the interaction parameters of the
effective theory depend on it? Finding the answer to this question is,
of course,
a very ambitious enterprise which will not be resolved in this paper.

One notable attempt, related to the approach of this paper, is the
work of L\"uscher, van Baal and collaborators \cite{baal} who use a
``non-perturbative'' approximation of Born-Oppenheimer type
wherein one tries to integrate out ``fast'' (short wavelength)
fluctuations thereby leaving an effective theory for the ``slow''
(long wavelength)
modes. In the case of QCD if this separation is carried out considering the
theory to be confined in a small box with periodic boundary conditions,
which acts as an infrared cutoff, then the fast modes can be integrated out
using perturbation theory. In fact, as the QCD vacuum is translationally
invariant, we can consider integrating out all non-constant modes thereby
yielding an effective field theory of constant gauge fields. In effect,
one is now considering a one-dimensional field theory, i.e.\ quantum mechanics.
In the case of the pure gauge theory the spectrum of the resulting Hamiltonian
will lead to the glueball spectrum. However, this is the glueball spectrum
associated with a ``universe'' that is smaller than a fermi. To describe the
real world it is necessary to take a box that has at least a few times the
size of the glueball.
Unfortunately, this leads to a breakdown in the perturbation theory that
is used to integrate out the non-constant modes. The physical reason
is simple: the perturbation theory is associated with free gluons, however,
the long wavelength modes are better described by other non-perturbative
effective degrees of freedom. Thus, although the methodology in principle
is a very powerful one, leading to an effective field theory that depends
only on the running coupling constant, in practice one finds that in
trying to extend it to larger boxes one encounters the same fundamental
problem that led to the introduction of the box in the first place.

Here we will take a somewhat less ambitious path motivating an ansatz for
an effective Hamiltonian via an analysis of the group-theoretical structure
of QCD, then analyzing some of the consequences of this ansatz, in particular
the spectrum of glueball masses. First of all we will take the effective
Hamiltonian to depend on spatially constant modes only. From the point
of view of the functional integral in principle it is always possible to
integrate out the non-constant modes.
However, if one is implementing an approximation technique
to do it one has to check that the resulting approximation is trustworthy. As
mentioned this has been investigated by L\"uscher, van Baal and collaborators
\cite{baal}. Their approximation breaks down when long wavelength gauge modes
(other than the zero mode) become strongly coupled. Nevertheless, given the
translation invariance of the QCD vacuum one would still expect an expansion
around the ground state to be an expansion around a spatially constant state.
We will take that attitude here and work with an effective Hamiltonian that
only depends on spatially constant modes.
We will restrict attention to the pure gauge sector of QCD. Hence, when
considering excited states we will only be considering glueballs. Given the
present experimental status, it is difficult to compare our results to
the outcome of actual measurements, however, we can compare with results
obtained from lattice simulations that restrict to the pure gauge
sector.

As a first step, in section 2, we formulate a classification scheme for a
many-gluon
system, assuming the dominance of only one mode with $J^P=1^-$.
This will enable us to order many-gluon states into bands defined by the
irreducible representations of an $O(8)$ group. This step alone plays
an important r\^ole, allowing us to explain the level ordering observed in
\cite{lattice1,peardon}. This result is to a large
extent model independent, the only assumption being that gluon
number (here we are referring to constituent gluons not perturbative
$g\rightarrow0$ gluons) is approximately a good quantum number, and
that the more gluons in a state the higher the corresponding energy.

Next, in section 3, we make an ansatz for the effective Hamiltonian of
QCD in terms of spatially constant gauge fields motivating the use
of a kinetic energy that is quadratic in momenta.
We further introduce a change of variables on the space of spatial gauge
potentials such that the new variables
are related to the intensity of the spatial gauge field, the spatial
quadrupole distribution and the rotational angles in space and colour space.
These variables have been widely used in nuclear physics \cite{dzublik,
complset}
and also in QCD \cite{baal,newvar}. Though we work with constant modes here,
this restriction is not necessary in general. In fact, the new variables as
functions of space and time qualify as
candidates for effective fields that better describe low energy QCD.
With a simple ansatz for the potential term the resulting
effective Hamiltonian turns out to be a function
of a one-dimensional oscillator, a five-dimensional oscillator and some
Casimir operators of symmetry groups appearing in the problem of many
gluon systems. The first oscillator
describes changes in the intensity of the gluonic field while the second
is related to the quadrupole deformation of the gluonic field with respect to
an ``intrinsic'' system yet to be defined (it will be similar in structure to
the ``body fixed frame'' of a nucleus). One feature of the effective
Hamiltonian is that there exists a conserved quantity which one can naturally
identify with the number of constituent gluons.

The resulting effective Hamiltonian depends on several parameters which we
determine in section 4 by fitting the corresponding spectrum to several
glueball states with spin, parity and charge
conjugation $J^{PC}$ and masses known from lattice gauge calculations.
We subsequently make predictions for all other states and compare, where
possible, with lattice simulations and results from other phenomenological
models, such as the MIT bag model. Finally in section 5 we draw our
conclusions.

\section{Gluon Taxonomy}

As a first step to understanding the structure of the spectrum of a many
gluon system one has to determine which values of $J^{PC}$ can appear,
where $J$ refers to spin and $P$, $C$ to parity and charge conjugation,
respectively. It is implicitly understood that only colour zero states are
allowed. To deal with this problem in an efficient way it is not sufficient
to consider the $SU(3)$ colour and $SO(3)$ spin groups only.\footnote{Note
that we use $SO(3)$ rather than $SU(2)$ as we are restricting our attention
to integer spins.} One reason for this is that gluon fields do
not transform according to the fundamental irreducible representation (irrep)
of the colour group. The main point, however, is that we are dealing with a
system of identical bosons, hence it is necessary to implement
Bose-Einstein statistics.
This can be done most conveniently by considering higher groups.
With respect to colour rotations, the gluonic fields transform
like  vectors in an eight-dimensional space so we can associate
a rotation group $O(8)$ with colour transformations.
Extending the latter rotations to unitary transformations, we
arrive at the group $U(8)$ \cite{hamermesh}. In
the same way the spatial rotations can be associated with an $U(3)$ group
containing the $SO(3)$ spin group. Bose-Einstein statistics are then
implemented by connecting at this level the spin and colour transformations
formally to an $U(24)$ group.

We thus obtain the following group chain \cite{viollier}:
\boldmath
\begin{equation}
\begin{array}{crcccl}
\mbox{\unboldmath$[N]$} & & & \mbox{\unboldmath$[h_1 h_2 h_3]$} & & \\
U(24) & \hspace{1cm}\supset\hspace{1cm} & U(8) & \times & U(3) & \\
 & \mbox{\unboldmath$\xi$} & \cup & & \cup & \\
 & \mbox{\unboldmath$(\omega_1 \omega_2 \omega_3 0)$} & O(8) & & SU(3)_J
& \mbox{\unboldmath$(p,q)$} \\
 & & \cup & & \cup & \mbox{\unboldmath$K$} \\
 & \mbox{\unboldmath$(0,0)$} & SU(3) & & SO(3) & \mbox{\unboldmath$J,M$}
\label{chain}
\end{array}
\end{equation}
\unboldmath
where for each group the quantum numbers of the corresponding irrep are given
and $\xi$ and $K$ are multiplicity indices appearing in the respective
reductions.
The symbol $[N]$ denotes the completely symmetric irrep of $U(24)$ for $N$
gluonic modes, which implements the Bose-Einstein statistics. The possible
irreps of $U(8)$ and $U(3)$ are then given by the Young tableaux
$[h_1 h_2 h_3]$, consisting of three rows with $h_1$, $h_2$ and $h_3$ boxes,
respectively, where $h_1 + h_2 + h_3 = N$. The fact that the $U(8)$ and $U(3)$
irreps are coupled to the completely symmetric irrep of $U(24)$ constrains
their Young tableaux to be equal \cite{hamermesh}.

The $U(3)$ group reduces via $SU(3)_J$,
where $J$ denotes spin, to the  rotation group $SO(3)$. The reduction rules
are given in Ref.\ \cite{elliott}, but will be summarized here for easier
reference. First the reduction from the irrep $[h_1 h_2 h_3]$ of $U(3)$
to $(p,q)$ of $SU(3)_J$ is simply given by $p = h_1 - h_2$ and $q = h_2 - h_3$.
The possible irreps $J$ (with magnetic quantum number $M$) of $SO(3)$ are
then determined by the possible values of the multiplicity index $K$,
\begin{equation}
K=\min(p,q), \min(p,q)-2, \ldots, 0 \;\,\mbox{or}\;\, 1 \;,
\label{multind}
\end{equation}
and
\begin{eqnarray}
J=\max(p,q), \max(p,q)-2, \ldots, 0 \;\,\mbox{or}\;\, 1 &\mbox{for}&
K=0 \nonumber \\
J=K, \ K+1, \ldots, K+\max(p,q) &\mbox{for}& K\ne 0 \:.
\label{elliott}
\end{eqnarray}

The group that describes rotations in colour space is $U(8)$, and it reduces
via $O(8)$ to the physical colour group $SU(3)$, where we restrict to colour
singlets, i.e.\ to the trivial irrep $(0,0)$. The intermediary orthogonal
group $O(8)$ will play a very important r\^ole in the following developments.
The $O(8)$ quantum numbers are $(\omega_1 \omega_2 \omega_3 0)$, where only
three of the four numbers can be non-zero as a consequence of the
restriction of the $U(8)$ irrep to three rows \cite{wyborne}. The largest
$O(8)$ irrep contained in a given irrep $[h_1 h_2 h_3]$ of $U(8)$ is the
one with $\omega_k=h_k$, $k = 1,2,3$. The precise rules for the reduction
of $U(8)$ to $O(8)$ are given, for instance, in Ref.\ \cite{wyborne} (there
is also a computer program available for arbitrary irreps of different
classical groups \cite{wyborne2}). The reduction of $O(8)$ to $SU(3)$ will be
carried out in a recursive manner using the reduction of $U(8)$ to
$O(8)$ just mentioned and the reduction of $U(8)$ to $SU(3)$. The
latter is given explicitly for up to five (constituent) gluons in Ref.\
\cite{viollier}, and the reduction for an arbitary irrep of $U(8)$ can be
found in Ref.\ \cite{ramon}. The recursive procedure for the reduction of
$O(8)$ to $SU(3)$ is described in detail in Appendix A. The results are
given for up to six constituent gluons in Table 1, considering only
those irreps of $U(8)$ and $O(8)$ which contain at least one colour zero
irrep. Finally, note that the multiplicities appearing in the
reduction of $U(8)$ to $O(8)$ and of $O(8)$ to $SU(3)$ are denoted globally
by $\xi$ in (\ref{chain}).

The above classification is not applicable in its present form to cases where
more than one gluonic mode is relevant. It is, however, always valid when one
particular --- even non-constant --- mode is
considered to be dominating at low energies, subject only to the condition
that it has spin 1, colour $(1,1)$ and, in view of our further considerations,
parity $(-1)$.

In order to discuss the properties of the states under parity and charge
conjugation we introduce boson creation and annihilation operators,
$b^{\,i\dagger}_a$ and $b^{\,i}_a$, for the gluon field $A^{\,i}_a$,
i.e.\ we write $A^{\,i}_a \sim (b^{\,i\dagger}_a + b^{\,i}_a )$ with a
proportionality factor
depending on certain parameters. What we will present in this section is
independent of these parameters, which will be fixed in the next section.
The operators $b^{\,i\dagger}_a$ and $b^{\,i}_a$ satisfy the usual
commutation relations.

A basis of states can be characterized unambiguously by the quantum
numbers in (\ref{chain}) (some of them are redundant, see also below).
Any such state can be obtained by applying the pair operators
\begin{equation}
{q^{[J]}_M}^\dagger = \sum_a [b^\dagger_a \times b^\dagger_a]^{[J]}_M
\label{qdagger}
\end{equation}
to a minimum weight state, i.e.\ a  state with $h_k=\omega_k$, $k = 1,2,3$.
The notation $[A\times B]^{[J]}_M$ in (\ref{qdagger}) represents the spin
coupling of two tensors $A$ and $B$ to spin $J$ \cite{edmonds}, where in our
case $J=0,2$ due to the spin and the bosonic nature of the gluons. Note also
that the pair operators $q^{[J]\dagger}_M$ are $O(8)$ scalars because of
the summation over $a$. The conjugate operators $q^{[J]}_M$ annihilate the
minimum weight states.\footnote{Together with the generators of the $U(8)$
group these operators form the algebra of the symplectic group $Sp(6,R)$,
which, however, will not be considered further here (see Ref.\ \cite{complset}
and references therein).}

A general basis state can then be decomposed as follows:
\begin{eqnarray}
& & \Big| [h_1 h_2 h_3] \, \xi \,
(\omega_1\omega_2\omega_3 0), K J M \Big\rangle \nonumber \\
&=& \left[ {\cal P}(q^\dagger)^{[2n_1,2n_2,2n_3]}\,\otimes\,
\Big| [\omega_1\omega_2\omega_3] \, \xi^\prime \,
(\omega_1\omega_2\omega_3 0) \Big\rangle \right] ^{[h_1 h_2 h_3] K J}_M \:.
\label{states}
\end{eqnarray}
The symbol ${\cal P}(q^\dagger)^{[2n_1,2n_2,2n_3] K_1 J_1}_{M_1}$ represents
a coupling of $(n_1+n_2+n_3)$ gluon pair operators from Eq.\ (\ref{qdagger})
to the $U(8)$ (or $U(3)$) irrep $[2n_1,2n_2,2n_3]$ and to spin $J_1$ with
multiplicity index $K_1$. Due to the commutativity of the pair operators
the Young diagram $[2n_1,2n_2,2n_3]$ must have an even number of boxes in
each row \cite{wyborne}. The operator ${\cal P}(q^\dagger)$ is subsequently
coupled with a minimum weight state to the total $U(8)$ (or $U(3)$) irrep
$[h_1 h_2 h_3]$ and to spin $J$, the $U(3)$ (or equivalently $U(8)$)
coupling being indicated by $\otimes$. This coupling is such that
\begin{equation}
N = h_1 + h_2 + h_3 = 2(n_1 + n_2 + n_3) + \omega_1 + \omega_2 + \omega_3 \:.
\label{constglue}
\end{equation}
It is natural to interpret the quantum number $N$ as the number of constituent
gluons. One should, however, be careful not to identify $N$ with the number
of the constituent gluons which arise in other phemonenological models, as
discussed in section 4. At any rate, Eq.\ (\ref{constglue}) gives the precise
definition of $N$ in our model.

We did not explicitly indicate in Eq.\ (\ref{states}) the quantum numbers $N$
and $(p,q)$, which are uniquely determined by $h_1, h_2, h_3$, the latter via
$p = h_1 - h_2$ and $q = h_2 - h_3$.
Furthermore it is understood that the state in (\ref{states}) is a singlet
with respect to the colour $SU(3)$.\footnote{In the pure gauge theory, the
gluons have to be coupled to colour $(0,0)$, which is not necessarily true
in the presence of quarks. The notation in Eq.\ (\ref{states}) can
be readily extended to the latter case.}
The multiplicity label $\xi$ is replaced by $\xi'$ in the second
representation in (\ref{states}) because part of the multiplicity is taken
care of by the different couplings
of $[2n_1,2n_2,2n_3]$ with $[\omega_1\omega_2\omega_3]$ to $[h_1h_2h_3]$.
The division of the state into a coupling of
$(n_1+n_2+n_3)$ gluon pairs and the ``rest'' is related to the concept
of generalized seniority, where a particle system can be divided into
pairs of particles coupled to colour zero and spin 0 or 2 and the rest
where no pairs of these types appear. The term ``generalized'' is used
because seniority is normally associated with a coupling to a total scalar.

The key point of the decomposition (\ref{states}) is that the gluon pair
operators change neither parity nor charge conjugation, which are therefore
already determined by the minimum weight states, i.e.\ the ones with
$[h_1 h_2 h_3] = [\omega_1\omega_2\omega_3]$. Furthermore, it turns out
that the minimum weight states carry definite parity and
charge conjugation, as will be demonstrated in the following. It is due to
these facts that the $O(8)$ irreps play a prominent r\^ole in
the classification of many-gluon states.

The parity of an $N$-gluon state is readily determined by noting that each
gluon field $A^{\,i}_a$ --- and therefore each operator $b^{\,i\dagger}_a$
--- carries parity $(-1)$. Therefore, the parity of a state with $O(8)$ irrep
$(\omega_1\omega_2\omega_3 0)$ is given by
\begin{equation}
P = (-1)^N = (-1)^{\omega_1+\omega_2+\omega_3} \:.
\label{parity}
\end{equation}

Obtaining the charge conjugation of the states is considerably more involved.
The reason can be seen by considering the simple example of a one-gluon state.
The properties of $A^{\,i}_a$ (and hence of $b^{\,i\dagger}_a$) under
charge conjugation are deduced in Ref.\ \cite{charge} and turn out to be
\begin{equation}
C A^{\,i}_a=\eta_a A^{\,i}_a \quad \mbox{(no sum over $a$)} \:,
\end{equation}
with
\begin{equation}
\eta_a = \left\{
\begin{array}{r@{\;\:\mbox{for}\;\:}l}
-1 & a=1,3,4,6,8 \\
+1 & a=2,5,7
\end{array} \right.
\label{eta}
\end{equation}
in the standard representation of $SU(3)$, where the generators are
proportional to the Gell-Mann matrices. Since one-gluon states belong
to the $U(8)$ and $O(8)$ irreps $[1]$ and $(1000)$, respectively, we have
an example of an $O(8)$ irrep lacking a definite charge conjugation.
As we will show in the sequel one can, contrary to this observation for
one-gluon states, associate definite charge conjugations with all colour-zero
states.

In order to see that, we have to construct the states from so-called
elementary permissible diagrams (epds) \cite{eps}. An epd is, in our case,
a minimum weight state as considered before, coupled to colour zero and
definite  spin, which is elementary in the sense that it cannot be decomposed
into other couplings. It will be denoted by
\begin{equation}
(N,[\omega_1\omega_2\omega_3],J) = [b^\dagger \otimes \cdots \otimes
b^\dagger]^{[\omega_1\omega_2\omega_3]J}_{M=J} \:,
\label{coupling}
\end{equation}
where $N=\omega_1+\omega_2+\omega_3$ and the $O(8)$ irrep is
$(\omega_1\omega_2\omega_3 0)$. Note that we always couple to $M=J$, which
implies that a product of two epds, say $(N, [h_1 h_2 h_3],J)
(N^\prime, [h^\prime_1 h^\prime_2 h^\prime_3],J^\prime)$,
represents a state with spin $J+J^\prime$. However, such a product of epds
has still to be projected onto a definite $U(3)$ (or equivalently $U(8)$)
irrep, meaning that the epds are defined here with respect to the maximum
weight of $SO(3)$ and not of $U(3)$ ($U(8)$).

Any minimum weight state can be represented by products of all
possible epds. Including the pair operators of Eq.\ (\ref{qdagger})
(with $M=J$) in the list of epds, we can then via (\ref{states}) construct
any state with the quantum numbers of (\ref{chain}) from the coupling of
epds.\footnote{As a detailed example for the construction of states with
the help of epds see the discussion of the five-dimensional harmonic
oscillator in Ref.\ \cite{sharp}.}
There exist systematic procedures to determine the total number of epds,
and the application of these general methods to the case at hand
(including states with colour which are important in the presence of quarks)
is presently
worked out \cite{epdsmir}. However, for the time being, we will adopt a more
pedestrian approach which is completely satisfactory in the case of small
gluon numbers. It should be noticed that in general the choice of the set of
epds is not unique being somewhat analogous to the choice of a basis for a
vector space.

The epds with up to three constituent gluons are (summation over repeated
indices is understood)
\begin{eqnarray}
(2,[2],J_1) & = & [b^\dagger_a \times b^\dagger_a]^{[J_1]}_{J_1}
\nonumber \\
(3,[3],J_2) & = & d_{abc}
\left[ [b^\dagger_a \times b^\dagger_b]^{[2]}\times b^\dagger_c
\right]^{[J_2]}_{J_2}
\nonumber \\
(3,[1^3],J_3) & = & f_{abc}
\left[ [b^\dagger_a \times b^\dagger_b]^{[1]}\times b^\dagger_c
\right]^{[J_3]}_{J_3} \:,
\label{epd1}
\end{eqnarray}
the first one being the pair operator (\ref{qdagger}), while the others
are minimum weight states in the sense defined before. The spins denoted by
$J_k$ are limited to the values $J_1=0,2$, $J_2=1,3$ and $J_3=0$. A complete
list of epds with up to six gluons, as well as the decomposition
of some minimum weight states with seven or eight gluons to be used
in the discussion of the glueball spectrum, can be found in Appendix B.

We can now use the decomposition of the states into epds to determine their
charge conjugation, using the fact that the epds are charge conjugation
eigenstates. For example, by construction the epd of two gluons in
(\ref{epd1}) has charge conjugation $C=+1$. For the epds with
three gluons we have for the symmetric coupling $C=-1$ and for the
antisymmetric coupling $C=+1$, which can be readily verified taking into
acount the combinations of $(a,b,c)$ which appear in these couplings and
using the list in (\ref{eta}). In general, as shown in Appendix C, the total
charge conjugation is simply given by
\begin{equation}
C=(-1)^{n_d} \:,
\end{equation}
where $n_d$ is the number of $d$-symbols appearing in the coupling.
Using this fact and the list of epds in Appendix B we can deduce the
charge conjugation of all states constructed via Eq.\ (\ref{states}) from
minimum weight states with up to six gluons. The results are given in
Table 1.

\section{An Effective Hamiltonian for QCD}

In this section we will motivate an ansatz for an effective Hamiltonian
for purely gluonic QCD at low energies. The Hamilton density in the temporal
gauge is of the form
\begin{equation}
{\cal H} = -\frac{1}{2}\sum_{i,a}(\partial_0 A^{\,i}_a)^2 + {\cal V}(A) \:,
\label{hamden}
\end{equation}
where the first term is the kinetic energy density and the second is the
potential which includes everthing else, such as
$(\nabla \times {\vec A}_a)^2$.

To obtain the effective action of the quantum theory
one must integrate over all possible physically
inequivalent configurations of the gauge field. This is, except in
certain special cases, impossible to do exactly. The most
popular approximation technique is perturbation theory, however, for a
discussion of QCD at low energies this simply will not do. Another
possible approximation, as mentioned previously, is of
Born-Oppenheimer type \cite{baal} wherein one tries to integrate out ``fast''
(short wavelength) fluctuations thereby leaving an effective theory for the
``slow'' (long wavelength)
modes. Here, we make an ansatz for the low-energy effective Hamiltonian,
taking it as a function of spatially constant gauge fields only.
Given the translation invariance of the QCD vacuum, this is not physically
unreasonable. The basic assumption behind this ansatz is that
the dynamics of the constant modes alone can give us information about the
glueball spectrum, i.e.\ the excited states of the theory, although obviously
the inner structure of the states cannot be described by constant modes.
As was shown in the work of L\"uscher, van Baal and Koller \cite{baal},
this assumption is correct for the system located inside
a box with periodic boundary conditions and a size of up to
0.7 fm (approximately).

After having integrated over all non-constant modes our general ansatz for the
quantized effective Hamiltonian is
\begin{equation}
H_{eff} = -\frac{1}{2B_\varrho}
\sum_{i,a}\frac{\partial^2}{(\partial {A^{\;i}_a})^2} + V_{eff}(A) \:,
\label{hamconst}
\end{equation}
where the effective potential $V_{eff}$ is a complicated function of
$A^{\,i}_a$ that will be modelled by a simple ansatz further below. The
approximation effected in (\ref{hamconst}) consists in neglecting higher
derivatives with respect to $A$, as well as terms mixing derivatives with
powers of $A$. This adiabatic approximation has been
justified in the case of a finite volume and to one-loop order
\cite{baal}, and here we will simply assume that for our purposes the same
approximation can also be applied to the more general case.
The constant $B_\varrho$ is a wave function renormalization,
which we will treat as an adjustable parameter.

After choosing the temporal gauge and integrating out all non-constant
modes there is still a residual gauge symmetry left in Eq.\
(\ref{hamconst}), namely the invariance under spatially constant, i.e.\
global (and time-independent) gauge transformations. We hence impose
on the physical states the condition
\begin{equation}
G_a \Psi_{phys}(A) = 0 \:, \label{nocolor}
\end{equation}
where the $G_a$ are the generators of global gauge transformations. Eq.\
(\ref{nocolor}) is equivalent to the statement that physical states have no
colour, in which form the condition was imposed on the many-gluon states
considered in the previous section. The consistency of the present gauge
fixing procedure with the integration over all non-constant modes was
shown in Ref.\ \cite{baal}.

We will now change variables. The transformation is chosen such that part of
the new variables is intimately related to the gluon pairs introduced in
the previous section. The interpretation of these variables will be given
below. Explicitly the transformation is given by \cite{complset}
\begin{equation}
A^{\,i}_a  = \sum_{k=1}^3 \varrho_k D^{1_3}_{k i}(\theta )
\Delta^{1_8}_{5+k,a} (\phi ) \:.
\label{trafo}
\end{equation}
where the constant modes of the vector potential $A^{\,i}_a$ depend on the
spatial indices
$(i=1,2,3)$ and the colour indices $(a=1,2,...,8)$.
$D^{1_3}_{k i}$ is the rotation matrix in the spatial
three-dimensional space with cartesian components and
$\Delta^{1_8}_{5+k,a}$ is the corresponding rotation matrix in the
colour space of dimension $8$. Of the latter only the last three
rows appear. The angles $\phi_{ab}$ for rotations in colour space can be
chosen in such a way that only 18 angles appear in Eq.\ (\ref{trafo}).
Together with the three angles $\theta_{ij}$ for spatial rotations and the
three $\varrho_k$ we then have (formally) 24 degrees of freedom as is the case
for the $A^{\,i}_a$ on the l.h.s.\ of Eq.\ (\ref{trafo}). Eight of the angles
$\phi_{ab}$ correspond to global gauge transformations, hence the
generators $G_a$ in Eq.\ (\ref{nocolor}) can be identified with the
angular momentum operators for the corresponding rotations. The condition
(\ref{nocolor}) then simply shows that physical states are independent
of these eight angles, thereby effectively reducing the number of degrees of
freedom to $2 \times 8$, as might have been anticipated.

Transformations similar to (\ref{trafo}) have been used in other areas of
physics \cite{dzublik,complset} and also in the work of van Baal and Koller
\cite{baal}. In both cases the
coordinates were taken to be space independent as in
Eq.\ (\ref{trafo}). However, Eq.\ (\ref{trafo}) can be readily extended to the
more general case where the coordinates include space (and time) dependence.
In any case, the interpretation of this coordinate transformation given below
is the same irrespective of whether the coordinates are space dependent or not.

We will now consider the interpretation of these new variables.  We first
define the composite field
\begin{eqnarray}
q_{ij} & = & \sum_a A^{\,i}_a A^{\,j}_a \nonumber \\
& = & \sum_{k} D^{1_3}_{k i}(\theta )
\varrho_{k}^2 D^{1_3}_{k j}(\theta ) \:.
\label{qij}
\end{eqnarray}
Eq.\ (\ref{qij}) gives the rotation in three-dimensional space of the
matrix $(q_{ij})$ into a system in which the matrix is diagonal, which will be
called the ``intrinsic system''.  The matrix
$(q_{ij})$ has both monopole and quadrupole components which are obtained
by simply coupling to the appropriate spin, i.e.\
\begin{equation}
q^{[J]}_M = \sum_{i,j} (1i,1j \mid JM) \, q_{ij} \:, \label{qlm}
\end{equation}
where $J=0,2$ indicates the spin and $(1i,1j \mid JM)$ is a Clebsch-Gordan
coefficient coupling two cartesian vectors to a spherical tensor
of spin $J$. Eq.\ (\ref{qlm}) represents of course the analogue to the gluon
pair operators $q^{[J]\dagger}_M$ of the previous section (see Eq.\
(\ref{qdagger})). For $J=0$,
\begin{equation}
q^{[0]}_0 = \frac{1}{\sqrt{3}}\sum_{i,a} (A^{\,i}_a)^2 =
\frac{1}{\sqrt{3}}\sum_k \varrho_k^2 \equiv \frac{\varrho^2}{\sqrt{3}}
\label{rhosquared}
\end{equation}
gives the square of the intensity of
the vector field. The spin $2$ part describes the quadrupole distribution
of the intensity of the vector field. Thus the $\varrho_k$ carry the
information about the monopole and quadrupole distribution.
In Eqs.\ (\ref{trafo}) and (\ref{qij}) the angles $\theta_{ij}$
give the orientation of the intrinsic system with respect to the laboratory
system and can be defined in many different ways. For instance, instead of the
rotation angles $\theta_{ij}$ in two-dimensional planes one could just use
Euler angles.
The angles $\phi_{ab}$ describe analogous rotations in colour space.

The monopole-quadrupole tensor $(q_{ij})$, being quadratic in the gauge field,
can only describe excitations with an even number of gluons. In addition, by
the summation over colour indices in (\ref{qij}), the variables and thus the
excitations they describe do not change parity nor charge conjugation.
Of course, ``non-pair'' excitations can still be described via (\ref{trafo})
and will be classified by $O(8)$ irreps.

In summary, our conjecture is that the variables $\varrho_k$, $\phi_{ab}$
and $\theta_{ij}$ represent effective degrees of freedom that will
be useful in describing the collective behaviour of the gluon field,
in the same way as in nuclear physics analogous variables describe
the collective rotational and vibrational modes of the nucleus.
We believe that a generalization of these variables to include
space-time dependence could also be useful in QCD.

As described in Ref.\ \cite{complset}, instead of the variables $\varrho_k$
one can use $\varrho$, $b$ and $c$ which are related to the former by
\begin{equation}
\varrho_k^2 = \frac{\varrho^2}{3} \left[ 1+2b\cos \left(c-\frac{2\pi}{3}k
\right) \right] \:, \label{robc}
\end{equation}
where $k=1,2,3$. The parameter $b$ gives the absolute deformation and $c$ its
deviation from axial symmetry. There are restrictions however
\cite{complset,greiner}: Firstly, the sector from $c=0$ to $\pi/3$
already represents all possible physically different
situations. The other sectors (six in total) can be mapped bijectively
to the one described \cite{greiner}. The Jacobian of the transformation
(\ref{robc}) becomes singular on the boundaries between different sectors.
Secondly, the restriction that the left-hand side of Eq.\ (\ref{robc}) be
positive imposes a $c$-dependent upper limit upon the variable $b$.

It will turn out to be even more convenient to replace $b$ by the variable
$\beta$ related to it via \cite{complset}
\begin{equation}
1+2 \left( \frac{\beta}{\sigma} \right)^2 =
\frac{1-b^2}{1-3b^2+2b^3 \cos(3c)}
\label{beta}
\end{equation}
with the scale factor $\sigma$ defined in Eq.\ (\ref{sig}) below.
The range of $\beta$ is from $0$ to $\infty$ (see Ref.\ \cite{greiner}).
For $b$ or $({\beta}/{\sigma})$ small we have ${\beta}/{\sigma} \approx b$.

Now we will rewrite the Hamiltonian of Eq.\ (\ref{hamconst})
in terms of the new variables of Eqs.\ (\ref{trafo})--(\ref{beta}).
For the time being we will concentrate on the kinetic term
$T = H_{eff} - V_{eff}$. In effect, for this case the transformation to the
new variables
has already been performed in Ref.\ \cite{complset} where the number $A$,
referring to the number of nucleons, has to be replaced by 9 and in
particular the orthogonal group $O(A-1)$ becomes $O(8)$, otherwise nothing
changes.

We will in the following
briefly outline the procedure and give the main definitions and results.
In the first step the kinetic term $T$ is rewritten in terms
of the variables $\varrho_k$ $(k=1,2,3)$ and the components
of the spin operator ${\bf J}_k$ in the intrinsic system associated
with the angles $\theta_{ij}$.
$T$ also depends on the generators ${\cal L}_{5+k_1,5+k_2}$
($k_1,k_2=1,2,3$) which form an $SO(3)$ subgroup of $O(8)$. As shown in
Refs.\ \cite{complset} and \cite{kinam}, the matrix elements of the
generators approach for large
\begin{equation}
\sigma^2 = \omega_1 + \omega_2 + \omega_3 + 6
\label{sig}
\end{equation}
the matrix elements of the generators ${\cal L}_k$ $(k=1,2,3)$
of the $SO(3)'$ subgroup of an $U(3)'$ group in the irrep
$[\omega_1\omega_2\omega_3]$. The primes indicate that a contraction
procedure \cite{gilmore} was used in terms of the expansion parameter
($1/\sigma^2$). The $[\omega_1\omega_2\omega_3]$ irrep of $U(3)'$ contains
the irrep $(\lambda =\omega_1 - \omega_2, \mu = \omega_2 - \omega_3)$ of its
$SU(3)'$ subgroup, and the latter can be reduced to $SO(3)'$ by using the
rules given in Eqs.\ (\ref{multind}) and (\ref{elliott}).
We will identify the components of ${\cal L}$ with angular momentum operators
later on.

The volume element in terms of the new variables is
\begin{equation}
(\varrho_1^2 - \varrho_2^2)(\varrho_1^2 - \varrho_3^2)
(\varrho_2^2 - \varrho_3^2)(\varrho_1\varrho_2\varrho_3)^{5}
d\varrho_1d\varrho_2d\varrho_3d\Omega_{\theta}d\Omega_{\phi} \:,
\label{volel1}
\end{equation}
where $d\Omega_\theta$ and $d\Omega_\phi$
refer to the volume element of the respective angles.
Their explicit form depends on the specific angles chosen (Euler or
others). We have also chosen $\varrho_1 \ge \varrho_2 \ge \varrho_3$
corresponding to the sector in the $(b,c)$-plane with
$0\le c \le \pi /3$ in order to avoid ambiguities in the choice of
the intrinsic system.

Next, we change the volume element via the transformation
\begin{equation}
\Phi  = (\varrho_1\varrho_2\varrho_3)^{5/2}
(\varrho_1^2+\varrho_2^2+\varrho_3^2)^2 \Psi
\label{volel2}
\end{equation}
to the new states $\Phi$ in the Schr\"odinger picture.
The change of volume element is such that for small
$\beta/\sigma \approx b$ the new volume element takes the form
\begin{equation}
\bigg( \frac{\beta}{\sigma} \bigg)^4 \sin(3c) \, d\varrho \, d\beta \, dc \,
d\Omega_{\theta} \, d\Omega_{\phi}
\label{volel3}
\end{equation}
as given in Ref.\ \cite{greiner}, neglecting corrections to the next
order in $(\beta/\sigma)$.

Implementing these changes, the kinetic term in the effective Hamiltonian
subsequently acquires the form
\begin{equation}
T = \frac{1}{2 B_\varrho} \left( - \frac{\partial^2}{\partial\varrho^2}
+ \frac{{\bf R^2}}{\varrho^2} \right) \:,
\label{kin}
\end{equation}
where ${\bf R}^2$ is an operator which contains linear and quadratic
derivatives with respect to the variables $\beta$ and $c$. It is also a
function of the spin operators ${\bf J}_k$ in the intrinsic system and
of the last three generators of the $O(8)$ group,
replaced by ${\cal L}_k$ after the contraction discussed above.
Additionally, it depends on combinations of Casimir operators
of the $O(8)$ group and of its subgroups $O(7)$, $O(6)$ and $O(5)$ in
the canonical chain.

In Ref.\ \cite{complset} an expansion is made in powers of the
parameter $(1/\sigma^2)$ defined in Eq.\ (\ref{sig}). In the case
of Ref.\ \cite{complset} this expansion parameter is always very small.
In the case at hand, where it starts from $1/6$ and decreases
with larger $\omega_k$ $(k=1,2,3)$, it is not too small, however one
would expect the expansion not to be too bad. Actually, we will expand not
only in powers of $(1/\sigma^2)$ but also in powers of $(\beta/\sigma)$,
leading to the contributions summarized by the formulas (5.6) to (5.11) of
Ref.\ \cite{complset}. However, it is necessary to keep the term given by
Eq.\ (5.11) in this reference (${\bf R}_2^2$ below) because it
can be, in our case, of the same order as $\sigma^2$.
We thus arrive at the following expression for the ${\bf R}^2$ operator
of Eq.\ (\ref{kin}):
\begin{equation}
{\bf R}^2 = {\bf R}_1^2 + {\bf R}_2^2 \:,
\end{equation}
where
\begin{eqnarray}
{\bf R}_1^2 & \approx &
2 \sigma^2 \Bigg[- \frac{1}{\beta^4}
\frac{\partial}{\partial \beta} \beta^4 \frac{\partial}{\partial \beta}
- \frac{1}{\beta^2 \sin (3c)}\frac{\partial}{\partial c}
\sin (3c) \frac{\partial}{\partial c} + \sum_{k=1}^3 \frac{{\bf L}_k^{2}}
{2\beta^2\sin^2 (c-2\pi k/3)} \Bigg] \nonumber \\
& & {}+ \Bigg(\sigma^4-\frac{9}{4}\Bigg)
\Bigg( 1+2 \bigg( \frac{\beta}{\sigma} \bigg)^2 \Bigg) + 12
\label{rlow0}
\end{eqnarray}
and
\begin{equation}
{\bf R}_2^2 \;\:\approx\;\: 2 \, {\cal C}_2(\lambda ,\mu) - 3
\sum_{k=1}^3 {\cal L}_k^2
\label{rlow}
\end{equation}
with
\begin{eqnarray}
{\bf L}_k & = & {\bf J}_k + {\cal L}_k \nonumber \\
{\cal C}_2(\lambda, \mu) & = & \lambda^2 + \lambda\mu + \mu^2
+3(\lambda + \mu ) \nonumber \\
\lambda & = & \omega_1 - \omega_2 , \quad \mu \;\:=\;\: \omega_2 -\omega_3
\:, \label{jc}
\end{eqnarray}
the $\omega_k$ being the quantum numbers of the $O(8)$ irrep.

The above is a good approximation in the case that $({\beta}/{\sigma})$ is
very small. However, as this is not entirely the case we will modify the
expression such that the contributions from higher orders in
$({\beta}/{\sigma})$ are taken into account via a redefinition
of the interaction parameters (a procedure commonly used in the collective
model of nuclei \cite{greiner}). At this point our model begins to look
even more phenomenological, similar to a Landau-Ginzburg
ansatz, where the theory contains parameters
which have to be adjusted to some
kind of ``experiment'', in our case to lattice calculations.
Nevertheless, the kinematical structure is maintained with the hope of
learning something about the spectral structure of pure gluonic QCD.

In order to illustrate the parameter redefinition we consider the term
in square brackets in Eq.\ (\ref{rlow0}). It can immediately be identified
with the kinetic energy associated with a quadrupole degree of freedom as
given in Ref.\ \cite{greiner}, where the spin operators ${\bf J}_k$ are
replaced by ${\bf L}_k = {\bf J}_k + {\cal L}_k$.

In principle, the full kinetic term (\ref{kin}) (after the reduction in
terms of the parameter $(1/\sigma^2)$) can be written down: it turns out
still to be quadratic in the derivatives, with the coefficients given as
power series in $({\beta}/{\sigma})$.
This is extensively discussed in Ref.\ \cite{greiner}.
It is straightforward but cumbersome to obtain the corresponding coefficients
of these higher terms in the kinetic energy.
A much simpler procedure takes the higher contributions into account by
redefining the coefficients of the lowest-order terms. In the example
considered above, we would multiply the square bracket in Eq.\ (\ref{rlow0})
by a parameter $(1/B_2)$, to be determined ``experimentally''. This procedure
is sensible as long as the expansion parameter
$({\beta}/{\sigma})$ does not vary over a wide range \cite{greiner}.

If we add to the terms in square brackets in Eq.\ (\ref{rlow0}) a
contribution proportional to $\beta^2$, we obtain the Hamiltonian of a
five-dimensional harmonic oscillator. In fact, such a term is available in
(\ref{rlow0}), and we multiply it --- following the above philosophy ---
by a factor $C_2$ (there is another term relatively suppressed by a factor
$(1/\sigma^4)$, which we neglect). We can then write the whole expression as
the number operator ${\bf N}_2$ of the quadrupole bosons times a constant
depending on $(C_2/B_2)$, plus the zero-point energy.

Taking into account all the terms in Eqs.\ (\ref{rlow0}) and (\ref{rlow}),
the operator ${\bf R}^2$ is given by a linear combination of the harmonic
oscillator of the quadrupole degree of freedom discussed above,
the eigenvalue ${\cal C}_2$ of the Casimir operator of the $SU(3)'$
group, the ${\cal L}^2$ operator of the $SO(3)'$ group contained in
$SU(3)'$, a term proportional to $\sigma^4$ and a constant.
Therefore, as an {\it ansatz} for ${\bf R}^2$ we propose
\begin{eqnarray}
{\bf R}^2 & = & 4 \sqrt{\frac{C_2}{B_2}}\sigma^2 {\bf N}_2
+ k_1 (\sigma^4-36) + k_2 {\cal C}_2(\lambda ,\mu )
+k_3 \sum_{k=1}^3 {\cal L}_k^2 + k_4 - \frac{1}{4} \:.
\label{eigenval1}
\end{eqnarray}
For convenience we have absorbed the zero-point energy of the harmonic
oscillator and a term $(36 k_1 + 1/4)$, in the parameter $k_4$ multiplying
the constant term.

Now we turn to the potential term $V_{eff}$ in (\ref{hamconst}). In general,
$V_{eff}$ will be a complicated function of the three fundamental
tensors $I_1=\varrho^2$, $I_2 =  \sqrt{1/5} \, \beta^2$ and
$I_3 =  -\sqrt{2/35} \, \beta^3 \cos (3c)$ (see Ref.\ \cite{greiner}), and
also of those angles $\phi_{ab}$ which are not excluded by gauge invariance.
In keeping with the expansion in powers of $(\beta/\sigma)$, we approximate
$V_{eff}$ by a function of $\varrho^2$ only. We also neglect any dependence
on the angles $\phi_{ab}$ which has to be considered as an {\em ad hoc} ansatz
in the present context. There is, however, some motivation for this omission
from the perturbative expressions calculated in Ref.\ \cite{baal}.
We will further approximate the potential in the vicinity of its minimum by a
linear combination of the terms $1/\varrho^2$ and $\varrho^2$, where the
former can be assimilated into ${\bf R}^2$ via a redefinition of the constant
$k_4$ in Eq.\ (\ref{eigenval1}). In fact, all the contributions to $V_{eff}$
which happen to have the same structure as one of the terms in ${\bf R}^2$,
can be taken into account by redefining the corresponding parameters.
The $\varrho^2$ term in the potential can be interpreted as a gluon mass
term as is clear from Eq.\ (\ref{rhosquared}). We remark that, despite this
mass term, the number of degrees of freedom of the spatially constant gluon
field corresponds to transverse gluons due to the remnant of Gauss' law
(\ref{nocolor}).

We then arrive at the final form of our ansatz for the effective Hamiltonian,
\begin{equation}
H_{eff} = -\frac{1}{2B_\varrho} \frac{\partial^2}{\partial\varrho^2}
+ \frac{{\bf R}^2}{2 B_\varrho\varrho^2}
+ \frac{C_\varrho}{2}
\varrho^2 \:,
\label{poteff}
\end{equation}
with ${\bf R}^2$ as in Eq.\ (\ref{eigenval1}) and a new parameter $C_\varrho$
from the modelling of the effective potential term. One of the main reasons
behind the specific approximations made for the kinetic and effective
potential terms is, of course, that the resulting Hamiltonian (\ref{poteff})
is separable and analytically solvable. ${\bf R}^2$ will be diagonalized
simultaneously with $H_{eff}$, hence we can replace ${\bf R}^2$
by its eigenvalue $R^2_\chi$, where $\chi$ refers to all quantum numbers
which appear in this eigenvalue. $H_{eff}$ then reduces to the Hamiltonian
of a harmonic oscillator in $\varrho$ with continuous angular momentum,
and the spectrum is
\begin{equation}
E^\chi_{n_\varrho} = \sqrt{\frac{C_\varrho}{B_\varrho}}
\Big( 2n_\varrho + \nu +4 \Big)
\label{envac}
\end{equation}
with $n_\varrho = 0,1,2,\ldots$ and
\begin{equation}
\nu = \sqrt{R^2_\chi + \frac{1}{4}} - 3 \:.
\label{lambda}
\end{equation}

Plugging in the eigenvalues for the different operators appearing in
${\bf R}^2$ in Eq.\ (\ref{eigenval1}), we obtain the more explicit expression
\begin{eqnarray}
E^{(\omega_1 \omega_2 \omega_3),{\cal L}}_{n_\varrho, N_2} & = &
\sqrt{\frac{C_\varrho}{B_\varrho}} \Bigg\{2 n_\varrho +
\Bigg[4 \sqrt{\frac{C_2}{B_2}} \sigma^2 N_2
+ k_1(\omega_1 + \omega_2 + \omega_3)(\omega_1 + \omega_2 + \omega_3 +12)
\nonumber \\
& & {}+ k_2 (\lambda^2 + \lambda\mu + \mu^2 + 3\lambda + 3\mu) +
k_3 {\cal L}({\cal L}+1) + k_4\Bigg]^{\frac{1}{2}}
- \sqrt{k_4}\Bigg\} \:,
\label{eigenval2}
\end{eqnarray}
where $N_2$ is the quantum number of the five-dimensional harmonic oscillator,
hence $N_2 = 0,1,2,\ldots$. As we are interested only in energy differences,
we have subtracted in (\ref{eigenval2}) the eigenvalue for the vacuum, which
has the quantum numbers $n_\varrho=0$, $N_2=0$ and $\omega_k=0$ ($k=1,2,3$),
consequently $\lambda = \mu = {\cal L} = 0$ from Eqs.\ (\ref{multind}),
(\ref{elliott}) and (\ref{jc}). We see explicitly from (\ref{eigenval2})
that $k_4$ is constrained to take non-negative values.

In our ansatz we have introduced a total of eight independent
parameters which have to be adjusted in order that $H_{eff}$ in
(\ref{poteff}) approximate as closely as possible the exact effective
Hamiltonian. Lacking knowledge of the latter, we will use information
from lattice simulations for the same purpose. In the next section,
we will determine the values of the six parameter combinations
\begin{equation}
\sqrt{\frac{C_\varrho}{B_\varrho}}, \: \sqrt{\frac{C_2}{B_2}}, \:
k_1, \: k_2, \: k_3, \: k_4 \:,
\label{parameters}
\end{equation}
appearing in Eq.\ (\ref{eigenval2}), by fitting the spectrum of $H_{eff}$ to
the glueball spectrum as measured in quenched lattice QCD.

Finally, we will give explicit expressions for the eigenstates of $H_{eff}$.
Following Ref.\ \cite{complset}, we have
\begin{eqnarray}
\lefteqn{ | \, n_\varrho N_2, \Lambda t L, J M, \Omega {\cal L},
(\omega_1 \omega_2 \omega_3) \delta \, \rangle \;\:=}
\nonumber \\
& & {\cal F}^\nu_{n_\varrho}(\varrho) F^\Lambda_{(N_2-\Lambda)/2}
(\beta) \sum_{ K^\prime} \Phi_{ K^\prime}^{\Lambda t L} (c)
\sum_{K,{\cal K}}(JK,{\cal L}{\cal K} \mid L K^\prime) D^J_{KM}(\theta)
D^{(\omega_1 \omega_2 \omega_3)}_{\Omega{\cal L} {\cal K} ; \delta (0,0)}
(\phi) \:. \label{complstate}
\end{eqnarray}
${\cal F}^\nu_{n_\varrho}$ are the eigenfunctions of the one-dimensional
harmonic oscillator in $\varrho$ given explicitly by
\begin{equation}
{\cal F}^\nu_{n_\varrho}(\varrho ) =
\left[ \frac{2 (B_\varrho C_\varrho )^{2+\nu/2} \, n_\varrho !}
{\Gamma ( n_\varrho + \nu + 4)} \right]^{\frac{1}{2}}
\varrho^{\nu +7/2}
L^{(\nu + 3)}_{n_\varrho} \left( \sqrt{B_\varrho C_\varrho} \,
\varrho^2 \right) e^{- \sqrt{B_\varrho C_\varrho} \, \varrho^2/2}
\:, \label{rostate}
\end{equation}
where $L^{(\alpha)}_n$ is a Laguerre polynomial and $\nu$ was defined in
(\ref{lambda}). The function $F$ in (\ref{complstate}) is the radial
part of the five-dimensional harmonic oscillator and $\Phi$ is the part
that depends on $c$. Both are given in Ref.\ \cite{sharp}.
$D^J_{KM}$ represents the Wigner D-function \cite{edmonds} whilst
the other D-function is the representation matrix of $O(8)$ in the
approximation of large $\sigma^2$. $\Lambda$ has the meaning
of seniority for the five-dimensional harmonic oscillator and gives the
number of $J=2$ gluon pairs not coupled to spin zero. The angular momentum
$L$ is restricted to the values allowed for a five-dimensional harmonic
oscillator \cite{sharp}, which in particular has no $L=1$ state.
The parameter $t$ is a
multiplicity index appearing in the classification of the five-dimensional
oscillator and $\Omega$ plays the r\^ole of the $K$ in Eqs.\ (\ref{multind})
and (\ref{elliott}) for the reduction of $SU(3)'$ to $SO(3)'$. Finally,
$\delta$ is a further multiplicity index in the reduction of $O(8)$ to $SU(3)$
colour.

The basis given in Eq.\ (\ref{complstate}) is one of the
possible realizations of the general expression given in Eq.\ (\ref{states}).
In principle the direct relation could be found by using Dragt's theorem
\cite{dragt} which relates the representation of states in terms of
boson creation operators to the coordinate representation. We can in
particular identify excitations of the one-dimensional oscillator (in
$\varrho$) and the five-dimensional oscillator (in $\beta$, $c$) with
gluon pairs created by the operators in (\ref{qdagger}). The relation is
given through Eqs.\ (\ref{qij})--(\ref{beta}) and involves
the parameters $B_\varrho, C_\varrho$ and $B_2, C_2$, respectively. The
total angular momentum of the pairs with quadrupole momentum is given
by $L$ (the other pairs do not contribute), where the allowed values of $L$
correspond to the fact that these gluon pairs represent identical bosons.
The minimum weight states in Eq.\ (\ref{states}) transforming according
to a certain $O(8)$ irrep $(\omega_1 \omega_2 \omega_3 0)$, are described
in (\ref{complstate}) by the corresponding $O(8)$ representation matrix.
We can interprete the value of ${\cal L}$ as the angular momentum of
these states, as is evident from considering a state with $L=0$,
where $J = {\cal L}$ by the Clebsch-Gordan coefficient in (\ref{complstate}).
In general, the total angular momentum $J$ of a state arises as the
(angular momentum) sum of $L$ stemming from the $J=2$ pairs and ${\cal L}$
originating from the minimum weight state. Interpreting the dependence
of the energy in (\ref{eigenval2}) on the quantum numbers of the states,
the tensor gluon pairs and the unpaired gluons (described by the minimum
weight states) interact strongly, while the scalar gluon pairs appear in
our approximation as free particles.

Using the basis (\ref{complstate}), we can calculate the expectation value
of $\varrho^2$ which turns out to be
\begin{equation}
\langle \varrho^2 \rangle = \frac{1}{C_\varrho}
E^\chi_{n_\varrho} \:,
\label{ro2}
\end{equation}
with $E^\chi_{n_\varrho}$ taken from (\ref{envac}), i.e.\ including the
vacuum energy. Eq.\ (\ref{ro2}) shows that the energy of a state is determined
exclusively by the square of the intensity of the gluon field. Let us
emphasize again that our model permits us only to calculate the spatial
field distribution for zero-momentum states of glueballs, therefore
we have no information about the internal structure of the states.

We are confident that the Hamiltonian presented in (\ref{poteff}) can serve
as a first approximation to the exact effective Hamiltonian. In future
developments the basis (\ref{complstate}) will then play the important r\^ole
of a perturbative basis in which to diagonalize the improved effective
Hamiltonian.
However, as there is at present no quantitative way to link our model to
fundamental pure QCD it cannot be regarded as more than an ansatz somewhat
in the spirit of a Landau-Ginzburg expansion of the free energy in
statistical physics. We would claim that it is more concretely
rooted in the fundamental microscopic theory than the former as
our order parameter
is directly related to the fundamental field content of QCD whereas,
for example in superconductivity, the Landau-Ginzburg ansatz does not
relate the scalar order parameter to the underlying
electron degrees of freedom.

\section{The Glueball Spectrum}

The values of the parameters in Eq.\ (\ref{eigenval2}) will be determined
by adjusting the spectrum of our model to the masses of the 16 glueball states
given by the lattice calculation of Ref.\ \cite{lattice1}.
For the three lowest-lying states we
use the values cited in Ref.\ \cite{michael} (rather than Ref.\
\cite{lattice1}) obtained as weighted
averages over several lattice measurements done by different groups
(see the references in
\cite{michael}). The $0^{--}$ glueball state comes with a huge error in
Ref.\ \cite{lattice1}, so we decided to take the value from Ref.\ \cite{luo}
instead, where a lattice Hamiltonian method was used. Finally, we
use some of the (preliminary) values of Ref.\ \cite{peardon} for the fit,
which result from a new generation of glueball measurements on anisotropic
lattices (see also Ref.\ \cite{morning}). We have employed the CERN-MINUIT
fitting routine \cite{cern} in order to obtain the best overall agreement.
The resulting fit is given by the following formula (cf. Eq.\
(\ref{eigenval2}))
\begin{eqnarray}
E^{(\omega_1 \omega_2 \omega_3),{\cal L}}_{n_\varrho, N_2} & = &
0.805 \, \mbox{GeV} \times \bigg\{ 2 n_\varrho +
\Big[ 1.17 {\sigma^2}N_2 + 0.165(\omega_1 + \omega_2 + \omega_3)(\omega_1 +
\omega_2 + \omega_3 + 12) \nonumber \\
& & {}+ 0.877 (\lambda^2 + \lambda\mu + \mu^2 + 3\lambda + 3\mu)
+0.028 {\cal L}({\cal L}+1) \Big]^{\frac{1}{2}} \bigg\} \hspace*{0.5cm} .
\label{energyfit}
\end{eqnarray}
Observe that the best value for the parameter $k_4$ is zero, so effectively
we have only used five parameters.
Note furthermore that the dependence on $J$ arises only indirectly
through $N_2$, ${\cal L}$ and the restriction given by the $SU(2)$
Clebsch-Gordan coefficients in Eq.\ (\ref{complstate}).

In Table 2 we list the results of the fit for the states considered in Ref.\
\cite{lattice1}.
For states marked with a cross only upper limits have been given in Ref.\
\cite{lattice1}. Before commenting on the quantitative aspects of the fits,
however, let us emphasize that {\em the overall ordering of glueball states
is to a large extent independent of the choice of the model Hamiltonian and
relies only on the assumption that the number of constituent gluons is
approximately a good quantum number and that more gluons correspond to higher
energy}. Of course, it is not clear a priori that the concept of a
constituent gluon is a sensible one. However, in our approximation the
quantity $N = h_1 + h_2 + h_3 = 2(n_\varrho + N_2) + \omega_1 + \omega_2
+ \omega_3$ is a good quantum number and can be interpreted as the number
of constituent gluons. It should be appreciated that the level ordering
of the glueball spectrum does not follow any easily recognizable pattern,
as was already stressed in Ref.\ \cite{lattice1}. Using the above ``minimal''
assumptions, however, we already obtain good agreement with the ordering
observed on the lattice. The two states that come out relatively too low,
$1^{--}$ and $3^{--}$, will be shifted upwards by large corrections
stemming from the Casimir ${\cal C}_2(\lambda,\mu)$ of the $SU(3)'$ subgroup
of the $U(3)'$ group related to $O(8)$ (discussed in the previous section).
Further below we will compare our results with those of other
effective models, such as the MIT bag or the flux-tube model.

Let us now proceed to a more quantitative comparison of our results with
the lattice measurements. It must be noted that the overall agreement is
surprisingly good, considering the simple-mindedness of our model. The
most significant deviation of our values from the results of Ref.\
\cite{lattice1} occurs for the $0^{+-}$ state. However, the statistical error
of the lattice measurement is probably too large for this discrepancy to
be taken seriously. Furthermore, our value is in excellent agreement with the
more recent calculation of Ref.\ \cite{peardon}. As far as the latter is
concerned, the statistical errors given in Ref.\ \cite{peardon} are
remarkably small, so if they are taken seriously, deviations from our values
are more clearcut in this case. There is, however, an uncertainty in the
overall energy scale for the measurements (not included in the
errors cited here), so there might be a proportionality factor involved when
going from one set of lattice measurements to another.
Comparing our results to those
of Ref.\ \cite{peardon} we find appreciable deviations for the $3^{+-}$
and $1^{--}$ states, but most importantly for the $0^{--}$ where the
deviation is twice as large as in all other cases. Our value is however in
excellent agreement with the central value given in Ref.\ \cite{luo} (after
normalizing to the mass of the $0^{++}$ according to Ref.\ \cite{michael}).

Of course, deviations from our results are to be expected, since our model
can only be considered as a first approximation to the exact effective
Hamiltonian obtained from QCD by integrating out all non-constant modes.
In particular, the form of the potential term (see the discussion preceding
Eq.\ (\ref{poteff})) is merely guessed at. As a consequence, one expects
the degeneracy of several states in our model to be lifted by the correponding
corrections. The good qualitative agreement with lattice results, however,
leads one to believe that those corrections are not so large as to render
our approximation meaningless.

In Table 3 we have gathered the predictions from our model for states
that may be accessible to future lattice measurements. Included are three
excited states that have already been measured on the lattice \cite{peardon},
the results being in good agreement with our predictions. On a (hyper)cubic
lattice one has to consider that rotational symmetry is
broken down to invariance under the cubic group $O$. There
exist only five irreps under spatial ``rotations'', termed $A_1, A_2, E,
T_1$ and $T_2$. Consequently, the identification of a state on the lattice
with a state in the continuum having angular momentum $J$ is ambiguous and
relies partially on the assumption that glueballs with higher $J$ are heavier
than those with smaller $J$, an assumption that is obviously not always
fulfilled. Tables listing the $O$ irreps contained in a given $SO(3)$ irrep
are available for example in Ref.\ \cite{bergb}.

State-of-the-art lattice calculations (on anisotropic lattices with
improved actions) are able to determine the energies of the ground states
and some first excited states for any given $O$ irrep \cite{peardon}.
Table 3 gives a list of our predictions for the energies of the first and
second excited states for all $R^{PC}$ ($R$ an irrep of $O$).
Due to the degeneracy of several states in our model, to be lifted by the
inclusion of corrections to our ansatz for the effective Hamiltonian,
the degree of excitation of a certain state (especially on the lattice)
cannot always be determined from our mass estimates.
In fact, these corrections
together with mass shifts due to possible mixing of different states with
the same $J^{PC}$ may change the sequence of states
substantially in energy regions with a high density of states (see Figures
2-4). Our predictions for the level ordering in these regions should be taken
with a grain of salt.

We would like to point out that within our model we obtain a $4^{++}$ state
lying considerably lower in mass than the $1^{++}$. Given that both $J=1$
and $J=4$ contain an irrep $T_1$ of the cubic group, on the lattice the
$4^{++}$ would have been mistaken, had it indeed been found, for a $1^{++}$
according to the above-mentioned rule of thumb.
Hence our value differs by more than two standard deviations from the
result of Ref.\ \cite{lattice1}, the latter being consistent with the more
recent calculation in Ref.\ \cite{peardon}.

The states that are marked by an asterisk in Table 3 are only accessible
on the lattice by measuring second excited states in the corresponding
$R^{PC}$ sector, a technical challenge which has as yet not been met.
More easily accessible states,
to be compared with our predictions, are $T_1^{-+*}, A_2^{+-*}, E^{+-*},
A_2^{--*}, E^{--*}$ and $T_1^{--*}$ corresponding to the continuum states
$4^{-+}, 3^{+-*}, 2^{+-*}, 3^{--*}, 2^{--*}$ and $1^{--*}$, respectively, with
the predicted massses given in Table 3. We should also mention that there are
particular problems with the mass determination of $0^{++}$ states on
anisotropic lattices, which is why we did not cite the corresponding results
in Tables 2 and 3. Ignoring these difficulties, the values given by the
collaboration of Refs.\ \cite{peardon,morning} are $(1.63 \pm .03)$ GeV for
the $0^{++}$ and $(2.84 \pm .04)$ GeV for the $0^{++*}$. While the first
value is consistent with the one given in Table 2 from Ref.\ \cite{michael},
the second one is in good agreement with our prediction.

In Fig.\ 2 we graphically compare our fit with the values for the glueball
masses as cited in Refs.\
\cite{lattice1,michael,luo}. The latter are represented by grey bars
including the one-sigma deviation for the statistical errors. Light grey
bars denote upper limits to the masses only. The other figures show our
predictions for the full glueball spectrum up to a certain energy, obtained
by using the parameters determined before. The states are ordered with
respect to their $PC$ eigenvalues and the $O(8)$ irreps.

The general prediction then is that the spectrum is {\em much denser} than
seen on the lattice, in particular in the sector $PC = ++$. Notice, however,
that there is no contradiction with the lattice results due to the fact
that on the lattice, with few exceptions, only the lowest-lying state in
a given representation $R^{PC}$ is measured. A potential problem is the
low-lying $4^{++}$ as discussed above.

Let us now discuss the present limits in energy for our prediction of
the spectrum, represented by the light grey bars in Figs.\ 2--4.
While the parity of any state is simply given by $P = (-1)^{\omega_1 +
\omega_2 + \omega_3}$ according to Eq.\ (\ref{parity}), we are
at present lacking a systematic determination of the charge conjugation
eigenvalues $C$ for states with more than six unpaired gluons, i.e.\
$\omega_1 + \omega_2 + \omega_3 > 6$ (there is work in progress on this
matter \cite{epdsmir}). We will consider first the sectors with positive
parity $P = +1$ and thus states with an even number of constituent gluons.
Above 5.08 Gev, there are possible states with at least eight unpaired
constituent gluons, the charge conjugation eigenvalues $C$ of which are as yet
undetermined. We do know, however, that the charge conjugation of states
with $O(8)$ irreps $(3320)$ and $(4220)$ is $C = -1$ (see Appendix B).
The lowest-lying of the $C$-undetermined states has
$(\omega_1 \omega_2 \omega_3 0) = (4330)$, i.e.\
10 unpaired constituent gluons, $(\lambda,\mu) = (1,0)$, ${\cal L} = 1$ and
$n_\varrho = N_2 = 0$, leading to $J^P = 1^+$ and a mass of 5.08 GeV. This
state sets the limit indicated in Figs.\ 2 and 3.

Turning now to the negative-parity sectors,
there are states with at least seven
unpaired constituent gluons, whose charge conjugation eigenvalues have yet to
be determined. The lowest-lying among them has $(\omega_1 \omega_2 \omega_3 0)
= (3310)$, $(\lambda,\mu) = (0,2)$, ${\cal L} = 0$ and $n_\varrho = N_2 = 0$,
and hence $J^P = 0^-$ and a mass of 4.46 GeV. All states with nine unpaired
constituent gluons have masses above that value. We have, however, determined
the charge conjugation of $(3220)$ to be $C = +1$ and that of $(4210)$ (not
indicated in the figure) to be $C = -1$. Using the latter fact one can show
that the lowest-lying $3^{-+}$ state has a mass of 5.03 Gev, although there
may be states with $PC = -+$ and other angular momenta between 4.46 and 5.03
GeV that we cannot classify completely at present. However, none of these
states can contain a representation $T_1$ or $A_2$ on the
lattice (i.e.\ $J \neq 1,3,4,\ldots$). Similarly, we can conclude that
the first state in this energy range containing $A_2^{--}$ is the $3^{--**}$
at 4.68 GeV. However, there are candidate states for $A_1^{--*}$,
$A_1^{--**}$ and $E^{--**}$ below this value, so for the time being we cannot
make predictions for these states on the lattice. All these facts have been
used in constructing Tables 2 and 3.

Finally, we would like to compare our results to those of other effective
models \cite{mitbag}--\cite{BS}. The oldest and also the most successful among
these is the MIT bag model \cite{mitbag}. All predictions for glueball
masses in this model that
we were able to find in the literature come out far too
low. If we, however, take the predictions of Donoghue, Johnson, Li (and of
Jaffe, Johnson for the $(TM)^2$ modes) \cite{mitbag} and simply scale the
energy up by a factor of about 2.2, the lattice results are roughly reproduced.
At this level, fine-structure corrections are absent, and the quality of
the approximation is comparable to our model when we just order the states
in accordance with the number of constituent gluons.

Multiplying the energies by a factor of 2.2 roughly corresponds to
changing $B^{1/4}$ to $2.2 \, B^{1/4}$, where $B$ is the bag constant that
characterizes the confining force. In the old bag model calculations this
constant has been adjusted to the meson spectrum. It then appears plausible
that $B^{1/4}$ has to be changed by a factor of $2.2 \approx 9/4$ in order to
describe glueballs, since the factor $9/4$ can be understood as the ratio of
the colour charge of gluons to that of quarks. Of course, the rather large
fine-structure corrections would have to be calculated anew to see if the MIT
bag model can really account for a quantitative description of the glueball
spectrum. As a first step, however, the masses of higher states should be
estimated in order to make sure that the part of the spectrum above
$\sim 4$ GeV as measured on the lattice can also be qualitatively understood
within this model. The calculations in Ref.\ \cite{viollier} can be taken as
a hint that one will have to consider states with more than three gluons.

Let us remark that the constituent gluons of the bag model are very different
from the ones in our model as they basically arise by putting free gluons
into the bag. Consequently, boundary conditions at the bag surface and
possible orbital motions play an important r\^ole. None of this makes any sense
in our effective Hamiltonian description. A potential advantage of the
bag model is that it allows, at least in principle, for a description of the
inner structure of the glueballs.

In Ref.\ \cite{operatordim} it was suggested to consider the canonical
dimension of the gauge-invariant operators that can create a glueball
with a given $J^{PC}$ out of the vacuum in order to obtain a qualitative
understanding of the level ordering of the spectrum. The results are
somewhat similar to the bag model (without fine structure corrections),
but compare less favourably with the lattice data. Of course, it is also
unclear how these ideas could be refined to yield quantitative predictions
for the spectrum.

There are other models like the ``glue-lego'' model \cite{lego},
the flux-tube model \cite{fluxtube} and the
non-relativistic potential model \cite{potential} which are, at least in
their original forms, more or less excluded by the available lattice data.
A relativistic version of the potential model is given by a recent
Bethe-Salpeter
calculation with massive constituent gluons \cite{BS}. With a gluon mass of
0.4 GeV very reasonable values for the masses of the $0^{++}, 2^{++}$ and
$0^{-+}$ glueballs are obtained, while the $2^{-+}$ comes out far too low and
may be spurious.
In Ref.\ \cite{swanson} a phenomenological QCD Hamiltonian
is combined with a BCS type vacuum, which leads to an effective mass for
the constituent gluons. Results are given for the masses of the $0^{++}$,
$2^{++}$, $0^{-+}$ and $2^{-+}$ glueballs including excited states. The
agreement with lattice data is satisfactory considering the approximations
made. We emphasize that again the constituent gluons in all these models are
different in nature from the ones in our model, and there is at present no
way to relate the two concepts.

Finally, sum rule calculations \cite{sumrules} have their place somewhere
between effective models and microscopic QCD. Up to now, predictions
have been obtained only for the three lowest-lying glueballs and these are
consistent with lattice measurements.

In some of the phenomenological models a $1^{-+}$ state with a relatively
low mass is predicted. This state is of some importance for experimental
glueball searches, since it cannot be obtained from the quark model for
mesons. In our model the $1^{-+}$ is a seven-gluon state and consequently
its mass is rather large, in agreement with the lattice data.
The problem with this state in other models can be traced to
additional longitudinal degrees of freedom for the massive gluons, which in
our case are absent as we have already pointed out in section 3.

We would like to conclude this section by remarking that in our model states
with $n_\varrho \ge 1$ (with the exception of the $0^{++}$ glueball) appear as
unbound states of $n_\varrho$ $0^{++}$ glueballs and the rest, insofar as the
total energy of the state is just the sum of the energies of these components.
Within the model, this is of course merely a consequence of the fact that the
$\varrho$-dependence of the effective Hamiltonian is given by a harmonic
oscillator. In physical terms, such states may correspond to very loosely
bound states of $n_\varrho$ $0^{++}$ glueballs and other glueballs. On the
lattice, these states may possibly be identified by analyzing their overlaps
with different lattice operators and finite-volume effects.
Such an enterprise is currently being envisaged \cite{peardon}.
{}From our model, there is precisely one candidate for such a weakly bound
state that has already been measured on the lattice, namely the $0^{-+*}$
state which should appear as a bound state of a $0^{++}$ glueball with a
$0^{-+}$ glueball. However, it must be noted that the lattice calculations
of Ref.\ \cite{peardon} give a larger mass for the $0^{-+}$ glueball and
a slightly smaller one for the $0^{-+*}$, so that the weak binding might be
merely an artefact of our approximation.

\section{Conclusions}

In this paper we have performed a group-theoretic analysis of many-gluon
states, based on the assumption that for the low-lying part of the
spectrum only one mode of the gluonic field (transforming under colour
rotations and parity transformations like a constant mode) is important.
The classification of the states involves a $U(8)$ colour group and a
$U(3)$ spin group. The former is reduced via an $O(8)$ group to the colour
group $SU(3)$. For each given parity and charge conjugation the irreps
of this $O(8)$ group classify bands of states. We remarked
that in the context of charge conjugation the decomposition into
elementary permissible diagrams (epds) proved very useful.
The systematic determination of epds for higher $O(8)$ irreps is presently
under investigation \cite{epdsmir}.

One of the outcomes of the analysis is the possibility of defining a certain
quantum number which we identified with the number of constituent gluons,
which again we emphasize are different from perturbative gluons
or constituent gluons arising in other phenomenological models.
The ordering of the states with respect to
this quantum number allows for a qualitative explanation of the level
ordering of glueball states with respect to spin as observed in lattice
calculations, assuming that the energy of the states increases
monotonically with the number
of constituent gluons. The ordering of the glueball states has been one
of the most puzzling results of lattice simulations. It is worth
noting that the MIT bag model might also be able to reproduce the glueball
spectrum, provided that the value of the bag constant is chosen differently
from the one used for quark states. To check this assertion new bag
model calculations for glueballs are needed.

In the second part of the paper we motivated a specific form for
an effective Hamiltonian of low-energy pure QCD which depended only
on the spatially constant modes of the gluon field.
We introduced collective coordinates related to the monopole and quadrupole
degrees of freedom of the intensity of the gluon field and rewrote the kinetic
part of the Hamiltonian in terms of these variables using the
results of Ref.\ \cite{complset}. Expanding in terms of $b$ and $(1/\sigma^2)$
where $b$ parametrizes the absolute deformation of the intensity distribution
and $\sigma^2$ is related to the quantum numbers of the $O(8)$ group,
and making an additional simplifying assumption for the potential term,
we were able to deduce to
lowest order a Hamiltonian which is a function of a one- and a five-dimensional
harmonic oscillator and of Casimir operators contained in the $U(8)\supset O(8)
\supset SU(3)$ chain of groups. Taking into account higher contributions via a
redefinition of the coefficients of these operators we arrived
at a phenomenological, QCD-motivated model Hamiltonian. The eigenstates
of this Hamiltonian could be classified by the quantum numbers
arising in the group-theoretical analysis, and in particular the number
of constituent gluons turned out to be a good quantum number.

By fitting the parameters appearing in the Hamiltonian we adjusted the
spectrum of the model Hamiltonian
to the ``experimental'' glueball spectrum, ``experimental'' in this context
meaning lattice simulations.
The general level of agreement was seen to be very good and predictions
for several further states could be made. In qualitative terms, we predicted
a much {\em denser} spectrum than that seen on the lattice without,
however, running into inconsistencies with present lattice measurements.
We have thus, with little input, principally the kinematical, i.e.\ group
theoretical, structure of QCD, been able to deduce a great deal
about the structure of an effective QCD Hamiltonian.

We can naturally envisage several possible further investigations.
Of course, an obvious challenge is the derivation of an effective
Hamiltonian from first principles. Without some form of infrared
cutoff, such as a fermi-sized box, this will require a solution of the
full crossover problem, i.e.\ a description of the interpolation between
quarks and gluons as effective degrees of freedom at high energies and
hadrons at low energies. The collective coordinates introduced
in this paper may prove useful in this context for the description of the
low-energy degrees of freedom. Furthermore, a theoretic
justification of the basic assumption made, that an effective Hamiltonian
depending only on the constant modes of the field can yield the full glueball
spectrum, is highly desirable. One could also think of possible extensions of
the model, for example by including composite operators, in order to describe
the inner structure of the physical states. Finally, it would of course be of
interest to try and define collective variables related to quarks and
anti-quarks, in a similar fashion to what has been done here for gluons.
In that way the model could be extended to describe quark-model mesons and
baryons, as well as the mixing between quarkic mesons and glueballs. Work on
these topics is in progress.

\subsection*{Acknowledgements}

P.O.H. thanks Yu.\ Smirnov for stimulating discussions concerning the
classification of many-gluon states and the DAAD
and DGAPA for financial
support. A.W. would like to thank C. Morningstar for communicating his
latest results for the glueball spectrum from anisotropic lattice
calculations. C.R.S. thanks P. van Baal for useful and interesting
discussions.
This research was partially funded by CONACyT grant number 3298P-E9607.

\begin{appendix}

\section*{Appendix}

\subsection*{\boldmath A \quad Reduction of $O(8)$ to $SU(3)$}

In this section will determine the reduction of the group $O(8)$ to $SU(3)$.
The reduction from $U(8)$ to $SU(3)$ is known \cite{viollier,ramon}
as is the reduction from $U(8)$ to $O(8)$ \cite{wyborne}. The reduction from
$O(8)$ to $SU(3)$ can then be obtained in a recursive manner, starting from
the simplest irrep $[0]$ in $U(8)$.

For the scalar irrep $[0]$ of $U(8)$, the reduction to $O(8)$ is $(0000)$
and the reduction to $SU(3)$
is $(0,0)$, hence the $SU(3)$-irrep $(0,0)$ is contained once in the
$(0000)$ irrep of $O(8)$, which is trivial.
Also the corresponding reduction of $U(3)$ via $SU_J(3)$
to $SO(3)$ is trivial, i.e. the $[0]$ irrep of $U(3)$
only contains the spin $J=0$. We have used the fact that the Young tableaux
of the $U(8)$ and $U(3)$ irreps have to be the same.

The $U(8)$ irrep of one
gluon (not presented in Table 1 because we give only those irreps of
$U(8)$ which contain at least one scalar irrep of $SU(3)$) reduces
to $(1000)$ of $O(8)$ \cite{wyborne} and to $(1,1)$ of $SU(3)$, consequently
the $(1000)$ irrep of $O(8)$ must contain the $(1,1)$ irrep of $SU(3)$.
For the corresponding reduction of $U(3)$ to $SO(3)$ we obtain $J=1$,
the expected result for a one-gluon state.

For the case of a two-gluon system we have the symmetric $[2]$ and
antisymmetric $[1^2]$ irreps of $U(8)$. The first reduces to $SU(3)$ as
$(2,2)+(1,1)+(0,0)$ (see Ref.\ \cite{viollier,ramon}), and for the
corresponding reduction of $U(3)$ to $SO(3)$ we find $J=0,2$, using the
rules of Eq.\ (\ref{elliott}). The reduction to
$O(8)$ for $[2]$ of $U(8)$ is given by
$(2000)+(0000)$ (Ref.\ \cite{wyborne}). The irrep $(0000)$
already appeared in the reduction of the $[0]$ irrep of $U(8)$, and we
know that it contains one colour zero irrep of $SU(3)$. The other
$SU(3)$ irreps appearing in the list of the reduction of $U(8)$ to $SU(3)$
are therefore contained in the $(2000)$ irrep of $O(8)$, and in particular
the latter cannot contain any colour scalar.
For the antisymmetric irrep $[1^2]$ of $U(8)$
the reduction to $SU(3)$ yields $(1,1)+(3,0)+(0,3)$ according to
Ref.\ \cite{viollier,ramon},
and the reduction to $O(8)$ is $(1100)$, hence $(1100)$ contains
precisely the $SU(3)$ irreps given above.

In this manner, we can proceed recursively towards higher $U(8)$ irreps,
thereby obtaining the complete reduction of $O(8)$ to $SU(3)$, and in
particular the number of colour singlets contained in a given $O(8)$ irrep.
The results for up to six constituent gluons are listed in Table 1, together
with the possible values of $J$ from the reduction to $SO(3)$.

\subsection*{B \quad Elementary Permissible Diagrams}

As claimed in section 2, any state characterized by the quantum numbers
in (\ref{chain}) can be obtained by a coupling of epds. Here we list
all the epds with up to six gluons:
\begin{eqnarray}
(2,[2],J_1) & = & [b^\dagger_a \times b^\dagger_a]^{[J_1]}_{J_1} \nonumber \\
(3,[3],J_2) & = & d_{abc}
\left[ [b^\dagger_a \times b^\dagger_b]^{[2]}\times b^\dagger_c
\right]^{[J_2]}_{J_2}
\nonumber \\
(3,[1^3],J_3) & = & f_{abc}
\left[ [b^\dagger_a \times b^\dagger_b]^{[1]}\times b^\dagger_c
\right]^{[J_3]}_{J_3}
\nonumber \\
(4,[2^2],J_4) & = & \hat{Y}_{[2^2]} * d_{abc}d_{ab^\prime c^\prime}
\left[ [b^\dagger_b \times b^\dagger_c]^{[2]} \times
[b^\dagger_{b^\prime} \times b^\dagger_{c^\prime}]^{[2]}\right]^{[J_4]}_{J_4}
\nonumber \\
(4,[2 1^2], J_5) & = & \hat{Y}_{[2,1^2]} * d_{abc}f_{ab^\prime c^\prime}
[b^\dagger_b \times b^\dagger_c]^{[0]}
[b^\dagger_{b^\prime} \times b^\dagger_{c^\prime}]^{[J_5]}_{J_5}
\nonumber \\
(5,[3 1^2],J_6) & = & \hat{Y}_{[3,1^2]} * f_{aa^\prime a^{\prime\prime}}f_{abc}
d_{a^\prime b^\prime c^\prime}
\left[ \left[ [b^\dagger_b \times b^\dagger_c]^{[1]}\times
[b^\dagger_{b^\prime} \times b^\dagger_{c^\prime}] \right]^{[J_c]} \times
b^\dagger_{a^{\prime\prime}} \right]^{[J_6]}_{J_6} \nonumber \\
(5,[2^2 1],J_7) & = & \hat{Y}_{[2^2,1]} *
f_{aa^\prime a^{\prime\prime}}f_{abc}
d_{a^\prime b^\prime c^\prime}
\left[ \left[ [b^\dagger_b \times b^\dagger_c]^{[1]}\times
[b^\dagger_{b^\prime} \times b^\dagger_{c^\prime}]\right]^{[J_c]} \times
b^\dagger_{a^{\prime\prime}}\right]^{[J_7]}_{J_7} \nonumber \\
(6,[3^2],J_8) & = & \hat{Y}_{[3^2]} *
d_{a_1a_2a_3}f_{a_1b_1c_1}f_{a_2b_2c_2}f_{a_3b_3c_3}
\Big[ [b^\dagger_{b_1} \times b^\dagger_{c_1}]^{[1]}  \nonumber \\
& & {}\times [b^\dagger_{b_2} \times b^\dagger_{c_2}]^{[J_b]}\times
[b^\dagger_{b_3} \times b^\dagger_{c_3}]^{[J_c]} \Big]^{[J_8]}_{J_8} \:.
\label{epd2}
\end{eqnarray}
On the right-hand side the Young operator $\hat{Y}$ appears,
which projects to a
definite symmetry with respect to $U(3)$ (or $U(8)$) and hence via
$[h_1h_2h_3]=$ $[\omega_1\omega_2\omega_3]$ to a definite $O(8)$ irrep.
The spins denoted by $J_k$ ($k=1,2,...,8$) are limited to the values
$J_1=0,2$, $J_2=1,3$, $J_3=0$, $J_4=0,2$, $J_5=1$,
$J_6 = 0,2$, $J_7=1$ and $J_8=1,3$. For the last three epds in
Eq.\ (\ref{epd2}) intermediate couplings with spin labels $J_b$ and $J_c$
appear, for each of which one of the possible values has to be chosen.
These are examples of the ambiguity in the choice of epds mentioned in
section 2.

The epds have been determined by a procedure described in
Appendix D using the reduction of $U(8)$ via $O(8)$ to $SU(3)$ as deduced
by the methods of Appendix A.
For the explicit construction of the epds we have used the fact that every
colour singlet state can be built up from the three ``fundamental tensors''
$\delta_{ab}$, $f_{abc}$ and $d_{abc}$ \cite{dittner}.
As a result, the decomposition
of all states with up to six consituent gluons in epds is known, and the
charge conjugations of the states can be obtained easily from the charge
conjugations of the corresponding epds. The latter are readily determined
from the explicit expressions (\ref{epd2}) as shown in Appendix C. The
results for the charge conjugations of all states with up to six gluons
are given in Table 1, together with their parity eigenvalues determined
by Eq.\ (\ref{parity}).

In order to reproduce the lowest $1^{-+}$ and $0^{+-}$ glueball states we
have to consider minimum weight states with seven and eight gluons,
respectively. The $1^{-+}$ state is contained in the $U(8)$
irrep $[32^2]$, which reduces to the $O(8)$ irreps
$(3220)$, $(3000)$, $(2200)$, and others that do not contain
colour singlets. Among the former, $(3000)$ and $(2200)$ can be obtained via
a product of already determined epds coupled to a definite $U(8)$ irrep.
The $(3220)$ irrep is given by a new epd,
\begin{eqnarray}
(7,[3 2^2],J=1) & = & \hat{Y}_{[3,2^2]} *
f_{aa_1a_2}d_{aa_3a_4}d_{a_1b_1c_1}d_{a_2b_2c_2}d_{a_3b_3c_3}
\bigg[ \Big[ [b^\dagger_{b_1} \times b^\dagger_{c_1}]^{[J_1]} \nonumber \\
& & {}\times [b^\dagger_{b_2} \times b^\dagger_{c_2}]^{[J_2]} \Big]^{[J_{12}]}
\times \Big[ [b^\dagger_{b_3} \times b^\dagger_{c_3}]^{[J_3]}\times
b^\dagger_{a_4} \Big]^{[J_{34}]} \bigg]^{[1]}_1 \:,
\label{1-+}
\end{eqnarray}
which has parity $P=-1$ and charge conjugation $C=+1$ (see Appendix C).

The $0^{+-}$ state can be obtained by multiplying the two epds $(3,[1^3],0)$
and $(5,[31^2],0)$ coupled to the $U(8)$ irrep $[42^2]$ and the $O(8)$
irrep $(4220)$. The charge conjugation of the state is the product of the
charge conjugations of each epd, hence $P=+1$ and $C=-1$.
In view of the later discussion of excited glueball states, we note that we
have determined in similar ways the
parity and charge conjugation of colour zero states in the $O(8)$ irrep
$(4210)$, contained in the $U(8)$ irrep $[421]$, to be $PC = --$, and
for the $O(8)$ irrep $(3320)$ contained in $[3^2 2]$ we have
obtained $PC = +-$.

\subsection*{C \quad Charge conjugation}

In order to determine the charge conjugation $C$ of a state contained in
a given $O(8)$ irrep $(\omega_1\omega_2\omega_3 0)$ (with $(0,0)$ as colour
for its $SU(3)$ subgroup) we introduced in section 2 the concept of epds.
A list of epds for up to six gluons is given in (\ref{epd2}). Having
determined the decomposition of a state into epds, its charge conjugation
can be obtained simply by multiplying the charge conjugations of the epds.

In order to deduce the properties of the epds under charge conjugation,
we use the results in Ref.\ \cite{charge}. Consider the monomial
\begin{equation}
T_{a_1a_2.....a_n} {b_{a_1}^{\,i_1}}^\dagger {b_{a_2}^{\,i_2}}^\dagger...
{b_{a_n}^{\,i_n}}^\dagger \:,
\label{mon1}
\end{equation}
where summation over the indices $a_k$ is understood.
Under charge conjugation the monomial transforms to
\begin{equation}
\eta_{a_1}\eta_{a_2}...\eta_{a_n}
T_{a_1a_2.....a_n} {b_{a_1}^{\,i_1}}^\dagger {b_{a_2}^{\,i_2}}^\dagger...
{b_{a_n}^{\,i_n}}^\dagger \:,
\label{mon}
\end{equation}
the values of the $\eta_a$ being given in Eq.\ (\ref{eta}).

As an example let us consider a special monomial of order four with colour
zero and a certain angular momentum coupling,
\begin{equation}
d_{abc}f_{ade}\left[ [b^\dagger_b\times b^\dagger_c]^{J_1}\times
[b^\dagger_d\times b^\dagger_e]^{J_2} \right]^{J}_M \:.
\label{mon2}
\end{equation}
Applying the charge conjugation operator and inserting a trivial factor
$\eta_a^2=1$, the $SU(3)$ coupling $d_{abc}f_{ade}$ transforms to
\begin{equation}
(\eta_a\eta_b\eta_cd_{abc})(\eta_a\eta_d\eta_ef_{ade}) \equiv
d^\prime_{abc}f^\prime_{ade} \:.
\label{etateta}
\end{equation}
Using Eq.\ (\ref{eta}) we note that
\begin{eqnarray}
d^\prime_{abc} & = & - d_{abc} \nonumber \\
f^\prime_{ade} & = &   f_{ade} \:,
\label{dfprime}
\end{eqnarray}
i.e.\ the $d$-symbol contributes a minus sign while the $f$-symbol
contributes a positive sign.

This can be readily extended to any monomial with colour zero.
All one has to do is count the number of $d$-symbols involved. Denoting this
number by $n_d$, the charge conjugation is given by
\begin{equation}
C = (-1)^{n_d} \:.
\label{chargec}
\end{equation}
The charge conjugations of the epds in (\ref{epd2}) follow immediately.

\subsection*{D \quad Construction of States}

In this appendix, we will describe how states with up to six constituent
gluons can be constructed from the epds listed in (\ref{epd2}). At the same
time, we will show how to actually determine the epds with the help of
the reduction of $U(8)$ via $O(8)$ to $SU(3)$ as given in Table 1.

We begin with the decomposition of a given state in a minimum weight state
and gluon pairs. As stated in section 2,
the lowest $U(8)$ irrep containing a given $O(8)$ irrep
$(\omega_1\omega_2\omega_3 0)$ is described by the Young tableau
$[\omega_1\omega_2\omega_3]$.
According to Eq.\ (\ref{states}) we can build up all higher $U(8)$ irreps
with the same $O(8)$ irrep by applying
an arbitary coupling of $(n_1+n_2+n_3)$ pair operators to the lowest state.
Each pair is represented by the $U(8)$ irrep $[2]$, and is identical with
the simplest epd $(2,[2],J_1)$ in the case $M_1=J_1$ (maximum weight with
respect to $SO(3)$).

A number of these pairs can be coupled to states represented by
Young diagrams with an
even number of boxes in each row. This restriction arises because the pairs
themselves represent identical bosons with six degrees of freedom each.
The problem of coupling of the pairs is
equivalent to the reduction \cite{wyborne}
\begin{eqnarray}
U(6) & \supset & U(3) \nonumber \\
{}[1] & \rightarrow & [2] \nonumber \\
{}[n] & \rightarrow & \sum \, [2n_1, 2n_2, 2n_3] \:,
\label{u6u3}
\end{eqnarray}
where in the general reduction in the last line of Eq.\ (\ref{u6u3}) the sum
is over different partitions of $n$, $n = n_1 + n_2 + n_3$.
For example, we have for the coupling of two and three pairs
\begin{eqnarray}
[2] \times [2] & = & [4] + [2^2] \nonumber \\
{}[2] \times [2] \times [2] & = & [6] + [42] + [2^3] \:,
\label{example}
\end{eqnarray}
respectively. The resulting irreps then have to be coupled with a minimum
weight state, $[h_1 h_2 h_3] = [\omega_1\omega_2\omega_3]$.

We will now construct all the states appearing in Table 1 in this manner.
Let us start by considering the $(0000)$ irrep of $O(8)$. Coupling with
one gluon pair operator, we obviously obtain the $U(8)$ irrep $[2]$.
The results of the coupling of two and three pairs are given in
Eq.\ (\ref{example}). We thus get all the states with up to six gluons
for the scalar irrep $(0000)$ of $O(8)$ (cf.\ Table 1).

In Table 1 we find two $U(8)$ irreps with three gluons, $[3]$ and $[1^3]$,
both of which cannot be constructed by a coupling of pair operators. They
hence give rise to epds, $(3,[3],J_2)$ and $(3,[1^3],J_3)$, which --- in
contradistiction to $(2,[2],J_1)$ --- are minimum weight states. We now
couple pair operators with these epds. As we exclusively consider states
with up
to six gluons, only one gluon pair will be coupled. We thus arrive at the
$U(8)$ irreps $[5]+[41]+[32]$ for the $O(8)$ irrep $(3000)$ and $[3 1^2]$
for $(1110)$ in agreement with Table 1.

Similarly, there appear states with four and five gluons in Table 1, namely
the $U(8)$ irreps $[2^2]$, $[2 1^2]$, $[2^2 1]$ and $[3 1^2]$, which
obviously cannot be obtained through coupling of the epds with two or three
gluons, and hence give rise to new epds themselves (see Eq.\ (\ref{epd2})).
We can still couple one gluon pair with the four-gluon epds to obtain the
states given in Table 1.

Finally, for the six-gluon states we consider couplings of the epds with
three gluons as for example $[3] \times [3] = [6] + [42]$, where similar
restrictions apply as for the coupling of several $[2]$ irreps of $U(8)$
(see Eq.\ (106) of Ref.\ \cite{wyborne}). We hence obtain new minimum weight
states for the $O(8)$ irreps $(6000)$ and $(4200)$. The other couplings
are $[3] \times [1^3] = [41^2]$ and $[1^3] \times [1^3] = [2^3]$, with
the corresponding minimum weight states $(4110)$ and $(2220)$.

There is one state in Table 1 which cannot be constructed in this way,
namely the $U(8)$ irrep $[3^2]$, which consequently yields the last epd
in (\ref{epd2}). The decomposition of the states in epds is useful because
it immediately determines the charge conjugation of the states in terms of
the charge conjugation of the epds (see the previous Appendix C).

\end{appendix}

\newpage
\section*{Table Captions}
\begin{description}
\item[Table 1:] %\hfill\\
Reduction of $U(8)$ to $O(8)$ up to six constituent
gluons taking into account only those $U(8)$ irreps which contain at least
one colour scalar (the actual number of colour scalars is indicated in the
third column). Also the
reduction of $U(3)$ via $SU_J(3)$ to the angular momentum group $SO(3)$ is
given, and we have indicated the corresponding values of $J$. Subindices
refer to multiplicities which can be distinguished by the multiplicity
index $K$. $P$ and $C$ refer to parity and charge conjugation,
respectively.

\item[Table 2:] %\hfill\\
Glueball masses calculated within the model compared to
lattice calculations. The values of the parameters used in our model are
given in Eq.\ (\ref{energyfit}). See sections 2 and 3 for our classification
of the glueball states displayed in the first four columns.
Columns five and six give the usual $J^{PC}$ classification along with the
value of the mass determined from our model Hamiltonian.

The lattice results which have been used to fit
the parameters are given in the last two columns. While the next to last
column cites the best confirmed values currently available, the last column
refers to preliminary results of a new generation of glueball calculations on
anisotropic lattices (we did not include the value for the $0^{++}$ glueball
due to problems with this specific state in the new method). The errors
displayed refer exclusively to statistical errors, while there might
be an additional overall variation in the scale. In particular, the scales
used in the last two columns may slightly differ from one another. Values
which are only upper limits, are marked by a cross. Some of the lattice
values have actually been read off the figures in the respective publications,
hence the data given might be erroneous in the last digit.

\item[Table 3:] %\hfill\\
Predictions from our model for states that can possibly be measured on the
lattice in the near future. The first five columns are as in Table 2,
while the sixth gives the corresponding representations on the lattice,
where the angular momentum $J$ is replaced by the label $R$ referring to
the representations of the cubic group $O$ (see Ref.\ \cite{bergb}). A
superscript asterisk marks an excited state in columns five and six (in
column six we have listed only those components which can be seen on the
lattice as at most second excitations). Where the degree of excitation of
a certain state could not be determined from our mass estimates
due to the degeneracy of several states in our model, possible higher degrees
of excitation were indicated by superscript asterisks in parenthesis.

The last column gives our mass estimates and a few preliminary results from
the lattice. An asterisk in front of a value indicates that the corresponding
state can only be determined from a second excitation on the lattice, which
appears to be rather demanding with present techniques.
\end{description}

\newpage
\section*{Figure Captions}

\begin{description}
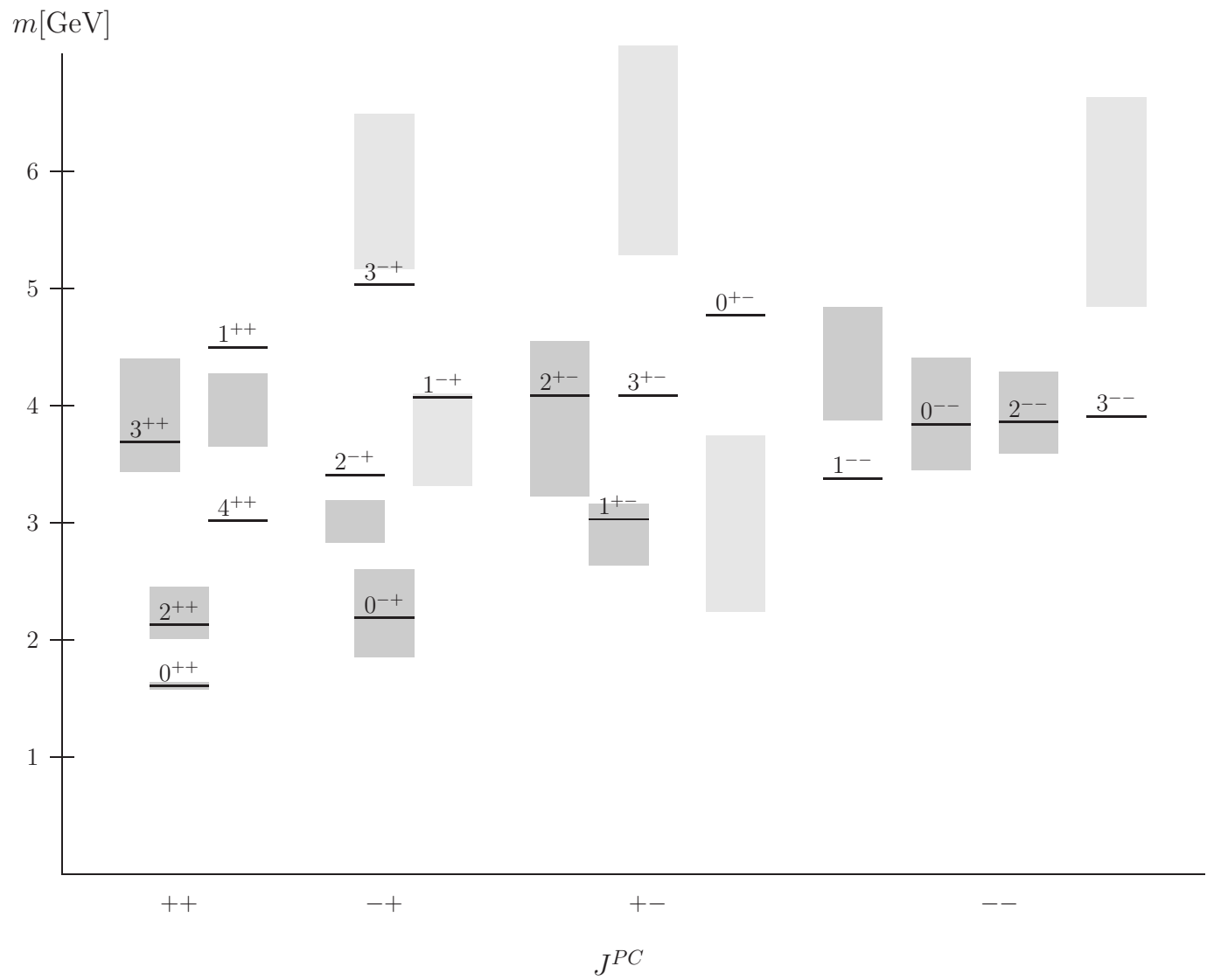
\item[Figure 1:] %\hfill\\
Comparison of our predictions and current ``best values'' for the glueball
masses as given in Table 2 (columns six and seven), arranged according
to their $PC$ eigenvalues. Our results are indicated
by solid lines, while the lattice values, including statistical errors
(one-sigma deviations), are represented by grey bars. Light grey bars
indicate upper limits. More than two-sigma deviations occur for the mass
of the $1^{++}$ versus our prediction of the $4^{++}$ (see the discussion in
section 4) and for the $0^{+-}$, where our
value is however in excellent agreement with the data of Ref.\ \cite{peardon}
(see Table 2).

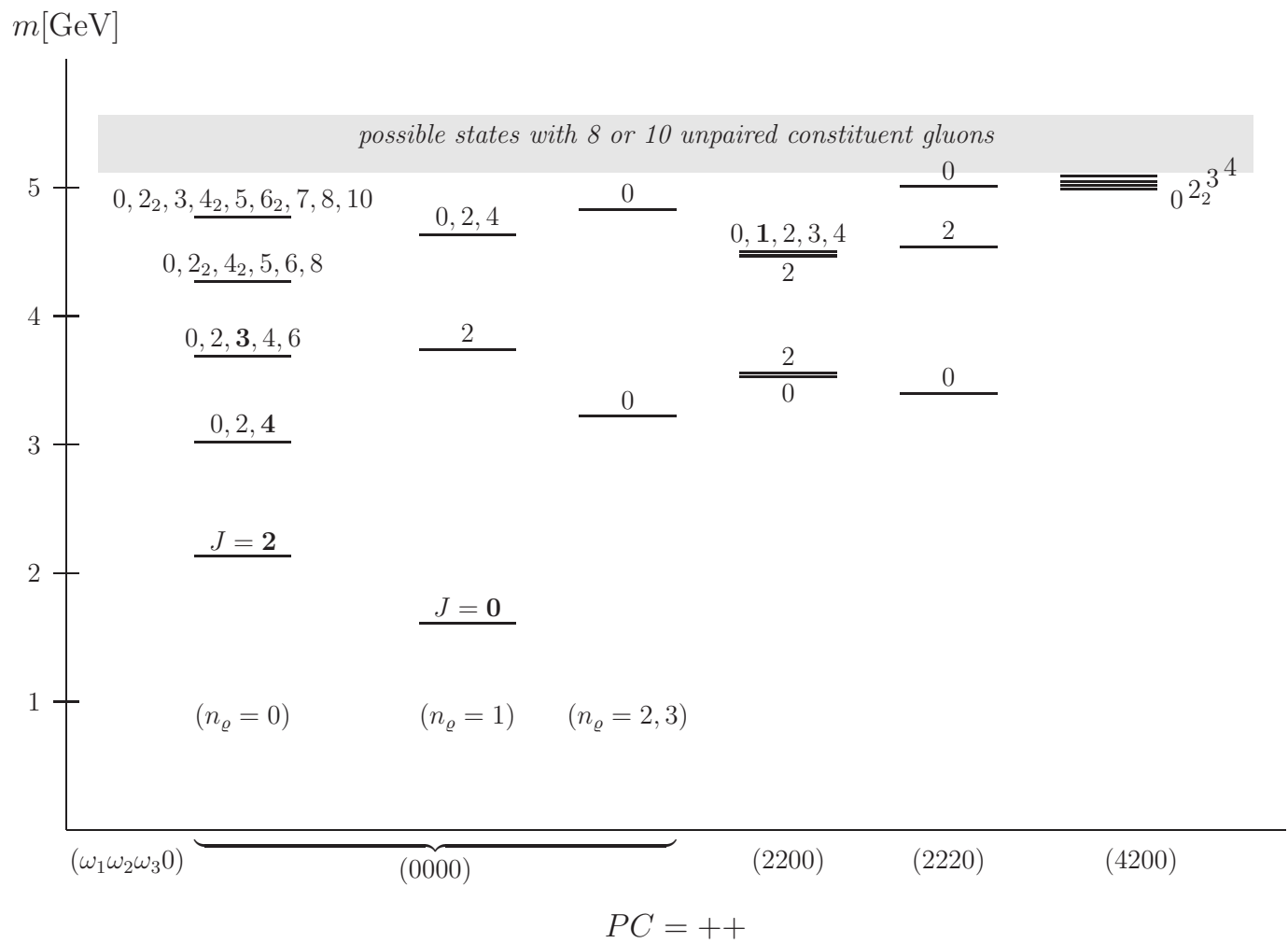
\item[Figure 2:] %\hfill\\
Our prediction for the complete glueball spectrum with $PC = ++$ below
5.08 GeV, using the parameters in (\ref{energyfit}). Note how dense the
spectrum is in comparison to what one would see on a lattice (Table 3).
States listed in Table 2 and Figure 1 are marked by bold face letters.
The lowest-lying state is, of course, the vacuum (not indicated in the
figure) which has $(\omega_1 \omega_2 \omega_3 0) = (0000)$, $(\lambda,\mu) =
(0,0)$, ${\cal L} = 0$ and $n_\varrho = N_2 = 0$ and hence $J^{PC} = 0^{++}$
and whose energy has been set to zero.

Above 5.08 Gev, there are possible states with at least 8 unpaired constituent
gluons ($\omega_1 + \omega_2 + \omega_3 \ge 8$). The charge conjugation
eigenvalues $C$ of these states are as yet unknown.

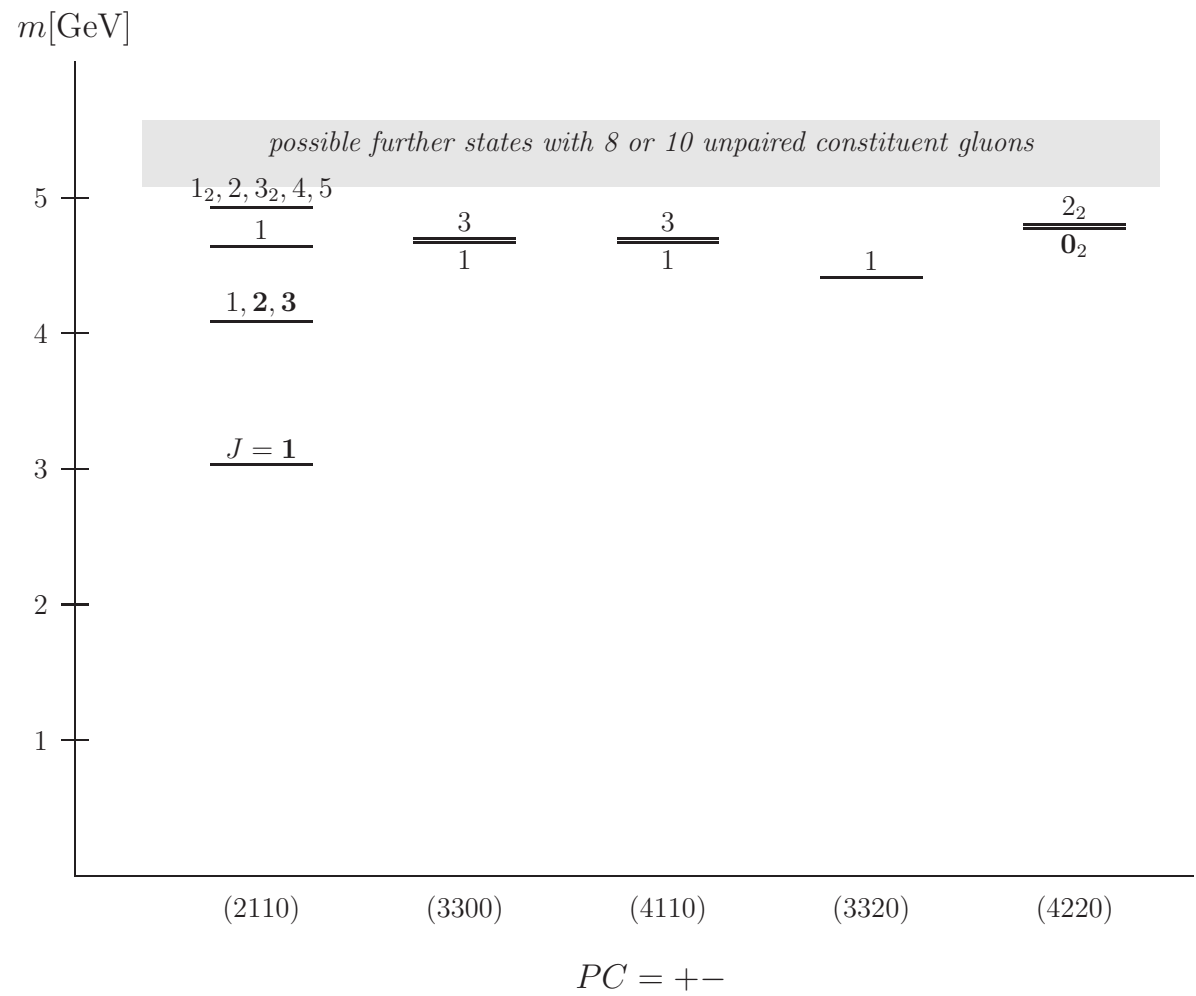
\item[Figure 3:] %\hfill\\
As in Figure 2, but for $PC = +-$. Note that the states with
$(\omega_1 \omega_2 \omega_3 0) = (3320)$ and $(4220)$ have 8 unpaired
constituent gluons. In these particular cases we have determined the charge
conjugation to be $C = -1$.

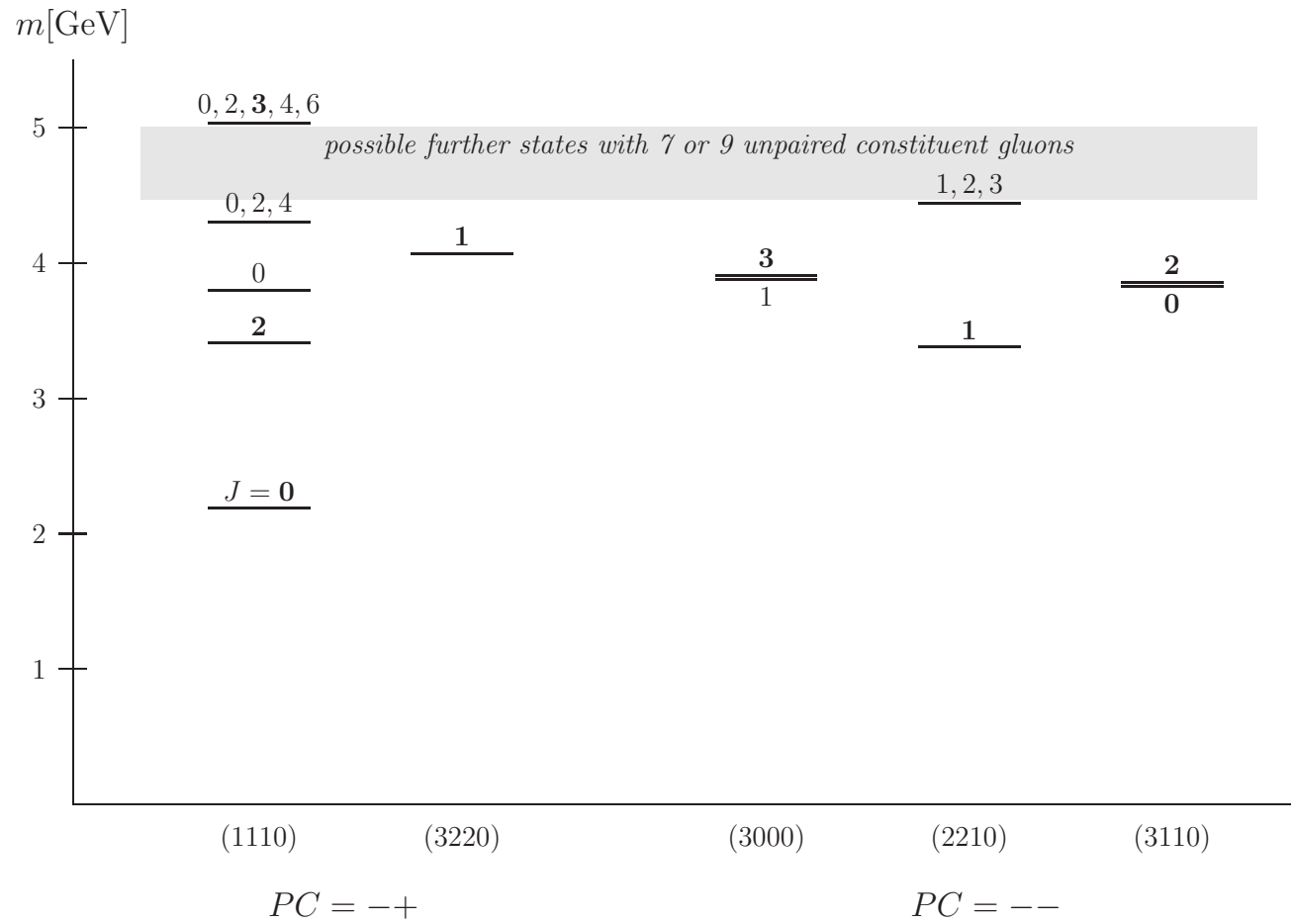
\item[Figure 4:] %\hfill\\
As in Figure 2, for states with $PC = -+$ and $--$ and masses below 4.46 GeV
(with the exception of the $(\omega_1 \omega_2 \omega_3 0) = (1110)$ states
at 5.03 GeV, see section 4). Between these two values, there are states with
at least 7 unpaired constituent gluons, whose charge conjugation eigenvalues
have yet to be determined. We do know, however, the charge conjugation of
states with $(\omega_1 \omega_2 \omega_3 0) = (3220)$ to be $C = +1$.
\end{description}

\newpage
\begin{table}
\begin{center}
\begin{tabular}{|c|c|c|c|c|c|}
\hline
$U(8)$ ($U(3)$) & $O(8)$ & \#(0,0) & SO(3) ($J$) & $P$ & $C$ \\
\hline
$[2]$ & (0000) & 1 & 0,2 & +1 & +1 \\
$[4]$ & (0000)& 1 & 0,2,4 & +1 & +1\\
$[2^2]$ & (0000) & 1 & 0,2 & +1 & +1 \\
$[6]$ & (0000) & 1 & 0,2,4,6 & +1 & +1 \\
$[4 2]$ & (0000) & 1 & 0,$2_2$,3,4 & +1 & +1 \\
$[2^3]$ & (0000) & 1 & 0 & +1 & +1 \\ \hline
$[3]$ & (3000) & 1 & 1,3 & $-1$ & $-1$ \\
$[5]$ & (3000) & 1 & 1,3,5 & $-1$ & $-1$ \\
$[4 1]$ & (3000) & 1 & 1,2,3,4 & $-1$ & $-1$ \\
$[3 2]$ & (3000) & 1 & 1,2,3 & $-1$ & $-1$ \\ \hline
$[1^3]$ & (1110) & 1 & 0 & $-1$ & +1 \\
$[3 1^2]$ & (1110) & 1 & 0,2 & $-1$ & +1 \\ \hline
$[2 1^2]$ & (2110) & 1 & 1 & +1 & $-1$ \\
$[4 1^2]$ & (2110) & 1 & 1,3 & +1 & $-1$ \\
$[3 2 1]$ & (2110) & 1 & 1,2 & +1 & $-1$ \\ \hline
$[2^2]$ & (2200) & 1 & 0,2 & +1 & +1 \\
$[4 2]$ & (2200) & 1 & 0,$2_2$,3,4 & +1 & +1 \\
$[3 2 1]$ & (2200) & 1 & 1,2 & +1 & +1 \\
$[2^3]$ & (2200) & 1 & 0 & +1 & +1 \\ \hline
$[2^2 1]$ & (2210) & 1 & 1 & $-1$ & $-1$ \\
$[3 1^2]$ & (3110) & 1 & 0,2 & $-1$ & $-1$ \\
$[6]$ & (6000) & 1 & 0,2,4,6 & +1 & +1 \\
$[4 2]$ & (4200) & 1 & 0,$2_2$,3,4 & +1 & +1 \\
$[4 1^2]$ & (4110) & 1 & 1,3 & +1 & $-1$ \\
$[3^2]$ & (3300) & 1 & 1,3 & +1 & $-1$ \\
$[2^3]$ & (2220) & 1 & 0 & +1 & +1 \\
\hline
\end{tabular}
\end{center}
\vspace{.5cm}
\centerline{\bf Table 1}
\end{table}

\newpage
\begin{table}
\begin{center}
\begin{tabular}{|c|c|c|c|c|c|c|c|}
\hline
$(\omega_1 \omega_2 \omega_3 0)$ & ${\cal L}$ $(\lambda , \mu)$
& $n_\varrho$ & $N_2$ & $J^{PC}$ & mass & lattice & Ref.\ \cite{peardon}
\\[-1.6mm]
& & & & & [GeV] & [Ref.] & (prelim.) \\
\hline
(0000) & 0 (0,0) & 1 & 0 & $0^{++}$ & 1.61 & 1.61$\pm$.03
\cite{michael} & --- \\
(0000) & 0 (0,0) & 0 & 1 & $2^{++}$ & 2.13 & 2.23$\pm$.22
\cite{michael} & 2.39$\pm$.01 \\
(0000) & 0 (0,0) & 0 & 3 & $3^{++}$ & 3.69 & 3.92$\pm$.48
\cite{lattice1} &   3.69$\pm$.04 \\
(2200) & 2 (0,2) & 0 & 1 & $1^{++}$ & 4.50 & 3.96$\pm$.31
\cite{lattice1} &  4.12$\pm$.05 \\ \hline
(1110) & 0 (0,0) & 0 & 0 & $0^{-+}$ & 2.19 & 2.23$\pm$.37
\cite{michael}&   2.59$\pm$.03 \\
(1110) & 0 (0,0) & 0 & 1 & $2^{-+}$ & 3.41 & 3.01$\pm$.18
\cite{lattice1} &  3.07$\pm$.02 \\
(3220) & 1 (1,0) & 0 & 0 & $1^{-+}$ & 4.07 & $\times$
3.71$\pm$.39 \cite{lattice1} &  4.18$\pm$.03 \\
(1110) & 0 (0,0) & 0 & 3 & $3^{-+}$ & 5.03 & $\times$
5.83$\pm$.66 \cite{lattice1} & 4.67$\pm$.05 \\ \hline
(2110) & 1 (1,0) & 0 & 0 & $1^{+-}$ & 3.03 & 2.90$\pm$.26
\cite{lattice1}&   2.94$\pm$.02 \\
(2110) & 1 (1,0) & 0 & 1 & $2^{+-}$ & 4.09 & 3.89$\pm$.66
\cite{lattice1} &   4.10$\pm$.04 \\
(2110) & 1 (1,0) & 0 & 1 & $3^{+-}$ & 4.09 & $\times$
6.18$\pm$.89 \cite{lattice1} &  3.54$\pm$.02 \\
(4220) & 0 (2,0) & 0 & 0 & $0^{+-}$ & 4.77 & $\times$
2.99$\pm$.75 \cite{lattice1} &  4.74$\pm$.05 \\ \hline
(2210) & 1 (0,1) & 0 & 0 & $1^{--}$ & 3.38  & 4.36$\pm$.48
\cite{lattice1} &  3.85$\pm$.04 \\
(3110) & 0 (2,0) & 0 & 0 & $0^{--}$ & 3.84 & 3.93$\pm$.48
\cite{luo} &  4.94$\pm$.05 \\
(3110) & 2 (2,0) & 0 & 0 & $2^{--}$ & 3.86 & 3.94$\pm$.35
\cite{lattice1} &  3.93$\pm$.02 \\
(3000) & 3 (3,0) & 0 & 0 & $3^{--}$ & 3.91 & $\times$ 5.74$\pm$.89
\cite{lattice1} & 4.13$\pm$.08 \\
\hline
\end{tabular}
\end{center}
\vspace{.5cm}
\centerline{\bf Table 2}
\end{table}

\newpage
\begin{table}
\begin{center}
\begin{tabular}{|c|c|c|c|c|c|c|}
\hline
$(\omega_1 \omega_2 \omega_3 0)$ & ${\cal L}$ $(\lambda , \mu)$
& $n_\varrho$ & $N_2$ & $J^{PC}$ & $R^{PC}$ & mass \\[-1.6mm]
& & & & & & [GeV] \\
\hline
(0000) & 0 (0,0) & 0 & 2 & $0^{++*}$ & $A_1^{++*(*)}$ & 3.02 \\
(0000) & 0 (0,0) & 0 & 2 & $2^{++*}$ & $E^{++*(*)},T_2^{++*(*)}$ & 3.02
\\[-1.6mm]
& & & & & & 3.29$\pm$.02 \cite{peardon} \\
(0000) & 0 (0,0) & 0 & 2 & $4^{++}$ & $A_1^{++*(*)},E^{++*(*)},$ & 3.02
\\[-1.6mm]
 & & & & & $T_1^{++},T_2^{++*(*)}$ & \\
(0000) & 0 (0,0) & 0 & 3 & $4^{++*}$ & $T_1^{++*(**)}$ & 3.69 \\
(0000) & 0 (0,0) & 0 & 3 & $6^{++}$ &  $A_2^{++(*)},T_1^{++*(**)}$ & 3.69 \\
(0000) & 0 (0,0) & 0 & 4 & $6^{++*}$ & $A_2^{++**}$ & $*$ 4.27 \\ \hline
(1110) & 0 (0,0) & 1 & 0 & $0^{-+*}$ & $A_1^{-+*}$ & 3.80 \\[-1.6mm]
& & & & & & 3.64$\pm$.04 \cite{peardon} \\
(1110) & 0 (0,0) & 0 & 2 & $0^{-+**}$ & $A_1^{-+**(*)}$ & $*$ 4.30 \\
(1110) & 0 (0,0) & 0 & 2 & $2^{-+*}$ & $E^{-+*(*)},T_2^{-+*(*)}$ &
4.30 \\[-1.6mm]
       &         &   &   &        & & 3.93$\pm$.02 \cite{peardon} \\
(1110) & 0 (0,0) & 0 & 2 & $4^{-+}$ & $A_1^{-+**(*)},E^{-+*(*)},$ &
4.30 \\[-1.6mm]
 & & & & & $T_1^{-+*},T_2^{-+*(*)}$ & \\
(1110) & 0 (0,0) & 0 & 3 & $4^{-+*}$ & $T_1^{-+**(**)}$ & $*$ 5.03 \\
(1110) & 0 (0,0) & 0 & 3 & $6^{-+}$ & $A_2^{-+(*)},T_1^{-+**(**)}$ &
5.03 \\ \hline
(2110) & 1 (1,0) & 0 & 1 & $1^{+-*}$ & $T_1^{+-*(*)}$ & 4.09 \\
(3300) & 3 (0,3) & 0 & 0 & $3^{+-*(*)}$ & $A_2^{+-*(*)},T_2^{+-**(*)}$ &
4.69 \\
(4110) & 3 (3,0) & 0 & 0 & $3^{+-*(*)}$ & $A_2^{+-*(*)},T_2^{+-**(*)}$ &
4.69 \\
(4220) & 0 (2,0) & 0 & 0 & $0^{+-*}$ & $A_1^{+-*}$ & 4.77 \\
(4220) & 2 (2,0) & 0 & 0 & $2^{+-*}$ & $E^{+-*}$ & 4.79 \\
(4220) & 2 (2,0) & 0 & 0 & $2^{+-**}$ & $E^{+-**}$ & $*$ 4.79 \\
(2110) & 1 (1,0) & 0 & 2 & $4^{+-}$ & $A_1^{+-**}$ & $*$ 4.93 \\ \hline
(3000) & 1 (3,0) & 0 & 0 & $1^{--*}$ & $T_1^{--*}$ & 3.88 \\
(2210) & 1 (0,1) & 0 & 1 & $2^{--*}$ & $E^{--*},T_2^{--**(*)}$ & 4.44 \\
(2210) & 1 (0,1) & 0 & 1 & $3^{--*}$ & $A_2^{--*},T_2^{--**(*)}$ & 4.44 \\
(3000) & 1 (3,0) & 0 & 1 & $3^{--**}$ & $A_2^{--**}$ & $*$ 4.68 \\
\hline
\end{tabular}
\end{center}
\vspace{.5cm}
\centerline{\bf Table 3}
\end{table}

\newpage
\begin{sidewaysfigure}
\begin{center}
\unitlength2cm
\begin{picture}(9.5,7.75)
%\put(0,0){\framebox(9.5,7.75){}}
\put(.5,1){\resizebox{17.5cm}{!}{
\begin{picture}(9.75,8)
\put(0,0){\line(1,0){9.75}}
\put(.5,-.5){\makebox(1,.5){$++$}}
\put(2.25,-.5){\makebox(1,.5){$-+$}}
\put(4.5,-.5){\makebox(1,.5){$+-$}}
\put(7.5,-.5){\makebox(1,.5){$--$}}
\put(3.75,-1.25){\makebox(2,1){\large $J^{PC}$}}
\put(0,0){\line(0,1){7}}
\put(-.5,0.5){\makebox(.5,1){$1$}}
\put(-.5,1.5){\makebox(.5,1){$2$}}
\put(-.5,2.5){\makebox(.5,1){$3$}}
\put(-.5,3.5){\makebox(.5,1){$4$}}
\put(-.5,4.5){\makebox(.5,1){$5$}}
\put(-.5,5.5){\makebox(.5,1){$6$}}
\put(-.1,1){\line(1,0){.2}}
\put(-.1,2){\line(1,0){.2}}
\put(-.1,3){\line(1,0){.2}}
\put(-.1,4){\line(1,0){.2}}
\put(-.1,5){\line(1,0){.2}}
\put(-.1,6){\line(1,0){.2}}
\put(-1,6.75){\makebox(2,1){\large $m$[GeV]}}
\thicklines
\put(.75,1.61){\colorbox{dark}{\rule[-.6mm]{0mm}{1.2mm}\hspace{1cm}}}
\put(.75,1.61){\makebox(.5,.25){$0^{++}$}}
\put(.75,1.61){\line(1,0){.5}}
\put(.75,2.23){\colorbox{dark}{\rule[-4.4mm]{0mm}{8.8mm}\hspace{1cm}}}
\put(.75,2.13){\makebox(.5,.25){$2^{++}$}}
\put(.75,2.13){\line(1,0){.5}}
\put(.5,3.92){\colorbox{dark}{\rule[-9.6mm]{0mm}{19.2mm}\hspace{1cm}}}
\put(.5,3.69){\makebox(.5,.25){$3^{++}$}}
\put(.5,3.69){\line(1,0){.5}}
\put(1.25,3.96){\colorbox{dark}{\rule[-6.2mm]{0mm}{12.4mm}\hspace{1cm}}}
\put(1.25,3.02){\makebox(.5,.25){$4^{++}$}}
\put(1.25,3.02){\line(1,0){.5}}
\put(1.25,4.5){\makebox(.5,.25){$1^{++}$}}
\put(1.25,4.5){\line(1,0){.5}}
\put(2.5,2.23){\colorbox{dark}{\rule[-7.4mm]{0mm}{14.8mm}\hspace{1cm}}}
\put(2.5,2.19){\makebox(.5,.25){$0^{-+}$}}
\put(2.5,2.19){\line(1,0){.5}}
\put(2.25,3.01){\colorbox{dark}{\rule[-3.6mm]{0mm}{7.2mm}\hspace{1cm}}}
\put(2.25,3.41){\makebox(.5,.25){$2^{-+}$}}
\put(2.25,3.41){\line(1,0){.5}}
\put(2.5,5.83){\colorbox{light}{\rule[-13.2mm]{0mm}{26.4mm}\hspace{1cm}}}
\put(2.5,5.03){\makebox(.5,.25){$3^{-+}$}}
\put(2.5,5.03){\line(1,0){.5}}
\put(3,3.71){\colorbox{light}{\rule[-7.8mm]{0mm}{15.6mm}\hspace{1cm}}}
\put(3,4.07){\makebox(.5,.25){$1^{-+}$}}
\put(3,4.07){\line(1,0){.5}}
\put(4.5,2.9){\colorbox{dark}{\rule[-5.2mm]{0mm}{10.4mm}\hspace{1cm}}}
\put(4.5,3.03){\makebox(.5,.25){$1^{+-}$}}
\put(4.5,3.03){\line(1,0){.5}}
\put(4,3.89){\colorbox{dark}{\rule[-13.2mm]{0mm}{26.4mm}\hspace{1cm}}}
\put(4,4.09){\makebox(.5,.25){$2^{+-}$}}
\put(4,4.09){\line(1,0){.5}}
\put(4.75,6.18){\colorbox{light}{\rule[-17.8mm]{0mm}{35.6mm}\hspace{1cm}}}
\put(4.75,4.09){\makebox(.5,.25){$3^{+-}$}}
\put(4.75,4.09){\line(1,0){.5}}
\put(5.5,2.99){\colorbox{light}{\rule[-15mm]{0mm}{30mm}\hspace{1cm}}}
\put(5.5,4.77){\makebox(.5,.25){$0^{+-}$}}
\put(5.5,4.77){\line(1,0){.5}}
\put(6.5,4.36){\colorbox{dark}{\rule[-9.6mm]{0mm}{19.2mm}\hspace{1cm}}}
\put(6.5,3.38){\makebox(.5,.25){$1^{--}$}}
\put(6.5,3.38){\line(1,0){.5}}
\put(7.25,3.93){\colorbox{dark}{\rule[-9.6mm]{0mm}{19.2mm}\hspace{1cm}}}
\put(7.25,3.84){\makebox(.5,.25){$0^{--}$}}
\put(7.25,3.84){\line(1,0){.5}}
\put(8,3.94){\colorbox{dark}{\rule[-7mm]{0mm}{14mm}\hspace{1cm}}}
\put(8,3.86){\makebox(.5,.25){$2^{--}$}}
\put(8,3.86){\line(1,0){.5}}
\put(8.75,5.74){\colorbox{light}{\rule[-17.8mm]{0mm}{35.6mm}\hspace{1cm}}}
\put(8.75,3.91){\makebox(.5,.25){$3^{--}$}}
\put(8.75,3.91){\line(1,0){.5}}
\end{picture}}}
\end{picture}
\end{center}
\vspace{.5cm}
\centerline{\bf Figure 1}
\end{sidewaysfigure}

\newpage
\begin{sidewaysfigure}
\begin{center}
\unitlength2cm
\begin{picture}(9.5,7)
%\put(0,0){\framebox(9.5,7){}}
\put(.5,1){\resizebox{17.5cm}{!}{
\begin{picture}(9.5,7)
\put(0,0){\line(1,0){9.5}}
\put(.1,-.5){\makebox(.75,.5){$(\omega_1 \omega_2 \omega_3 0)$}}
\put(1,-.75){\makebox(3.75,1){$\underbrace{\hspace{7.5cm}}_{\D (0000)}$}}
\put(5.25,-.5){\makebox(.75,.5){$(2200)$}}
\put(6.5,-.5){\makebox(.75,.5){$(2220)$}}
\put(8,-.5){\makebox(.75,.5){$(4200)$}}
\put(3.75,-1.25){\makebox(2,1){\large $PC = ++$}}
\put(0,0){\line(0,1){6}}
\put(-.5,0.5){\makebox(.5,1){$1$}}
\put(-.5,1.5){\makebox(.5,1){$2$}}
\put(-.5,2.5){\makebox(.5,1){$3$}}
\put(-.5,3.5){\makebox(.5,1){$4$}}
\put(-.5,4.5){\makebox(.5,1){$5$}}
\put(-.1,1){\line(1,0){.2}}
\put(-.1,2){\line(1,0){.2}}
\put(-.1,3){\line(1,0){.2}}
\put(-.1,4){\line(1,0){.2}}
\put(-.1,5){\line(1,0){.2}}
\put(-1,5.75){\makebox(2,1){\large $m$[GeV]}}
\thicklines
\put(.25,5.12){\colorbox{light}{\rule{0mm}{8.9mm}\hspace{18cm}}}
\put(.25,5.15){\makebox(9,.5){\em possible states with 8 or 10
unpaired constituent gluons}}
\put(1,.75){\makebox(.75,.25){$(n_\varrho = 0)$}}
\put(1,2.13){\makebox(.75,.25){$J = \bf 2$}}
\put(1,2.13){\line(1,0){.75}}
\put(1,3.02){\makebox(.75,.25){$0,2,\bf 4$}}
\put(1,3.02){\line(1,0){.75}}
\put(1,3.69){\makebox(.75,.25){$0,2,{\bf 3},4,6$}}
\put(1,3.69){\line(1,0){.75}}
\put(1,4.27){\makebox(.75,.25){$0,2_2,4_2,5,6,8$}}
\put(1,4.27){\line(1,0){.75}}
\put(1,4.77){\makebox(.75,.25){$0,2_2,3,4_2,5,6_2,7,8,10$}}
\put(1,4.77){\line(1,0){.75}}
\put(2.75,.75){\makebox(.75,.25){$(n_\varrho = 1)$}}
\put(2.75,1.61){\makebox(.75,.25){$J = \bf 0$}}
\put(2.75,1.61){\line(1,0){.75}}
\put(2.75,3.74){\makebox(.75,.25){$2$}}
\put(2.75,3.74){\line(1,0){.75}}
\put(2.75,4.63){\makebox(.75,.25){$0,2,4$}}
\put(2.75,4.63){\line(1,0){.75}}
\put(4,.75){\makebox(.75,.25){$(n_\varrho = 2,3)$}}
\put(4,3.22){\makebox(.75,.25){$0$}}
\put(4,3.22){\line(1,0){.75}}
\put(4,4.83){\makebox(.75,.25){$0$}}
\put(4,4.83){\line(1,0){.75}}
\put(5.25,3.28){\makebox(.75,.25){$0$}}
\put(5.25,3.53){\line(1,0){.75}}
\put(5.25,3.56){\makebox(.75,.25){$2$}}
\put(5.25,3.56){\line(1,0){.75}}
\put(5.25,4.22){\makebox(.75,.25){$2$}}
\put(5.25,4.47){\line(1,0){.75}}
\put(5.25,4.5){\makebox(.75,.25){$0,{\bf 1},2,3,4$}}
\put(5.25,4.5){\line(1,0){.75}}
\put(6.5,3.4){\makebox(.75,.25){$0$}}
\put(6.5,3.4){\line(1,0){.75}}
\put(6.5,4.54){\makebox(.75,.25){$2$}}
\put(6.5,4.54){\line(1,0){.75}}
\put(6.5,5.01){\makebox(.75,.25){$0$}}
\put(6.5,5.01){\line(1,0){.75}}
\put(8.5,4.91){\makebox(.75,.25){$0^{\D \, 2_2^{\D \, 3^{\D \, 4}}}$}}
\put(7.75,4.99){\line(1,0){.75}}
\put(7.75,5.02){\line(1,0){.75}}
\put(7.75,5.05){\line(1,0){.75}}
\put(7.75,5.09){\line(1,0){.75}}
\end{picture}}}
\end{picture}
\end{center}
\vspace{.5cm}
\centerline{\bf Figure 2}
\end{sidewaysfigure}

\newpage
\begin{sidewaysfigure}
\begin{center}
\unitlength2cm
\begin{picture}(8.25,7)
%\put(0,0){\framebox(8.25,7){}}
\put(.5,1){\resizebox{15cm}{!}{
\begin{picture}(8.25,7)
\put(0,0){\line(1,0){8.25}}
\put(1,-.5){\makebox(.75,.5){$(2110)$}}
\put(2.5,-.5){\makebox(.75,.5){$(3300)$}}
\put(4,-.5){\makebox(.75,.5){$(4110)$}}
\put(5.5,-.5){\makebox(.75,.5){$(3320)$}}
\put(7,-.5){\makebox(.75,.5){$(4220)$}}
\put(3.25,-1.25){\makebox(2,1){\large $PC = +-$}}
\put(0,0){\line(0,1){6}}
\put(-.5,0.5){\makebox(.5,1){$1$}}
\put(-.5,1.5){\makebox(.5,1){$2$}}
\put(-.5,2.5){\makebox(.5,1){$3$}}
\put(-.5,3.5){\makebox(.5,1){$4$}}
\put(-.5,4.5){\makebox(.5,1){$5$}}
\put(-.1,1){\line(1,0){.2}}
\put(-.1,2){\line(1,0){.2}}
\put(-.1,3){\line(1,0){.2}}
\put(-.1,4){\line(1,0){.2}}
\put(-.1,5){\line(1,0){.2}}
\put(-1,5.75){\makebox(2,1){\large $m$[GeV]}}
\thicklines
\put(.5,5.08){\colorbox{light}{\rule{0mm}{9.7mm}\hspace{15cm}}}
\put(.5,5.15){\makebox(7.5,.5){\em possible further states with 8 or 10
unpaired constituent gluons}}
\put(1,3.03){\makebox(.75,.25){$J = \bf 1$}}
\put(1,3.03){\line(1,0){.75}}
\put(1,4.09){\makebox(.75,.25){$1,{\bf 2},\bf 3$}}
\put(1,4.09){\line(1,0){.75}}
\put(1,4.64){\makebox(.75,.25){$1$}}
\put(1,4.64){\line(1,0){.75}}
\put(1,4.93){\makebox(.75,.25){$1_2,2,3_2,4,5$}}
\put(1,4.93){\line(1,0){.75}}
\put(2.5,4.42){\makebox(.75,.25){$1$}}
\put(2.5,4.67){\line(1,0){.75}}
\put(2.5,4.7){\makebox(.75,.25){$3$}}
\put(2.5,4.7){\line(1,0){.75}}
\put(4,4.42){\makebox(.75,.25){$1$}}
\put(4,4.67){\line(1,0){.75}}
\put(4,4.7){\makebox(.75,.25){$3$}}
\put(4,4.7){\line(1,0){.75}}
\put(5.5,4.41){\makebox(.75,.25){$1$}}
\put(5.5,4.41){\line(1,0){.75}}
\put(7,4.52){\makebox(.75,.25){${\bf 0}_2$}}
\put(7,4.77){\line(1,0){.75}}
\put(7,4.8){\makebox(.75,.25){$2_2$}}
\put(7,4.8){\line(1,0){.75}}
\end{picture}}}
\end{picture}
\end{center}
\vspace{.5cm}
\centerline{\bf Figure 3}
\end{sidewaysfigure}

\newpage
\begin{sidewaysfigure}
\begin{center}
\unitlength2cm
\begin{picture}(9,6.5)
%\put(0,0){\framebox(9,6.5){}}
\put(.5,1){\resizebox{16.5cm}{!}{
\begin{picture}(9,6.5)
\put(0,0){\line(1,0){9}}
\put(1,-.5){\makebox(.75,.5){$(1110)$}}
\put(2.5,-.5){\makebox(.75,.5){$(3220)$}}
\put(4.75,-.5){\makebox(.75,.5){$(3000)$}}
\put(6.25,-.5){\makebox(.75,.5){$(2210)$}}
\put(7.75,-.5){\makebox(.75,.5){$(3110)$}}
\put(1,-1.25){\makebox(2,1){\large $PC = -+$}}
\put(5.75,-1.25){\makebox(2,1){\large $PC = --$}}
\put(0,0){\line(0,1){5.5}}
\put(-.5,0.5){\makebox(.5,1){$1$}}
\put(-.5,1.5){\makebox(.5,1){$2$}}
\put(-.5,2.5){\makebox(.5,1){$3$}}
\put(-.5,3.5){\makebox(.5,1){$4$}}
\put(-.5,4.5){\makebox(.5,1){$5$}}
\put(-.1,1){\line(1,0){.2}}
\put(-.1,2){\line(1,0){.2}}
\put(-.1,3){\line(1,0){.2}}
\put(-.1,4){\line(1,0){.2}}
\put(-.1,5){\line(1,0){.2}}
\put(-1,5.25){\makebox(2,1){\large $m$[GeV]}}
\thicklines
\put(.5,4.47){\colorbox{light}{\rule{0mm}{10.6mm}\hspace{16.5cm}}}
\put(.5,4.61){\makebox(8.25,.5){\em possible further states with 7 or 9
unpaired constituent gluons}}
\put(1,2.19){\makebox(.75,.25){$J = \bf 0$}}
\put(1,2.19){\line(1,0){.75}}
\put(1,3.41){\makebox(.75,.25){$\bf 2$}}
\put(1,3.41){\line(1,0){.75}}
\put(1,3.8){\makebox(.75,.25){$0$}}
\put(1,3.8){\line(1,0){.75}}
\put(1,4.3){\makebox(.75,.25){$0,2,4$}}
\put(1,4.3){\line(1,0){.75}}
\put(1,5.03){\makebox(.75,.25){$0,2,{\bf 3},4,6$}}
\put(1,5.03){\line(1,0){.75}}
\put(2.5,4.07){\makebox(.75,.25){$\bf 1$}}
\put(2.5,4.07){\line(1,0){.75}}
\put(4.75,3.63){\makebox(.75,.25){$1$}}
\put(4.75,3.88){\line(1,0){.75}}
\put(4.75,3.91){\makebox(.75,.25){$\bf 3$}}
\put(4.75,3.91){\line(1,0){.75}}
\put(6.25,3.38){\makebox(.75,.25){$\bf 1$}}
\put(6.25,3.38){\line(1,0){.75}}
\put(6.25,4.44){\makebox(.75,.25){$1,2,3$}}
\put(6.25,4.44){\line(1,0){.75}}
\put(7.75,3.58){\makebox(.75,.25){$\bf 0$}}
\put(7.75,3.83){\line(1,0){.75}}
\put(7.75,3.86){\makebox(.75,.25){$\bf 2$}}
\put(7.75,3.86){\line(1,0){.75}}
\end{picture}}}
\end{picture}
\end{center}
\vspace{.5cm}
\centerline{\bf Figure 4}
\end{sidewaysfigure}


\begin{thebibliography}{99}
\bibitem{mitbag}  A. Chodos, R. L. Jaffe, K. Johnson, C. B. Thorn and
V. F. Weisskopf, Phys.\ Rev.\ {\bf D9} (1974), 3471 \\
R. L. Jaffe and K. Johnson, Phys.\ Lett.\ {\bf B60} (1976), 201 \\
J. F. Donoghue, K. Johnson and B. A. Li, Phys.\ Lett.\ {\bf B99} (1981), 416 \\
T. Barnes, F. E. Close and S. Monaghan, Phys.\ Lett.\ {\bf B110} (1982), 159;
Nucl.\ Phys.\ {\bf B198} (1982), 380 \\
C. E. Carlson, T. H. Hansson and C. Peterson, Phys.\ Rev.\ {\bf D27} (1983),
1556; {\em ibid.}\ {\bf D30} (1984), 1594 \\
M. Chanowitz and S. Sharpe, Nucl.\ Phys.\ {\bf B222} (1983), 211
\bibitem{lego} D. Robson, Nucl.\ Phys.\ {\bf B130} (1977), 328
\bibitem{fluxtube} N. Isgur and J. Paton, Phys.\ Lett.\ {\bf B124} (1983),
247; Phys.\ Rev.\ {\bf D31} (1985), 2910
\bibitem{potential} J. M. Cornwall and A. Soni, Phys.\ Lett.\ {\bf B120}
(1983), 431 \\
W.-S. Hou and A. Soni, Phys.\ Rev.\ {\bf D29} (1984), 101
\bibitem{BS} J. Y. Cui, J. M. Wu and H. Y. Jin, Phys.\ Lett.\ {\bf B424}
(1998), 381
\bibitem{swanson} A. Szczepaniak, E. S. Swanson, C.-R. Ji and S. R. Cotanch,
Phys.\ Rev.\ Lett.\ {\bf 76} (1996), 2011
\bibitem{efftheo} For a review and references see, for example, J. F. Donoghue,
E. Golowich and B. R. Holstein, ``Dynamics of the Standard Model'', Cambridge
Monographs on Particle Physics, Nuclear Physics and Cosmology No.\ 2,
Cambridge University Press (1992)
\bibitem{baal} M. L\"uscher, Phys.\ Lett.\ {\bf B118} (1982), 391; Nucl.\
Phys.\ {\bf B219} (1983), 233 \\
M. L\"uscher and G. M\"unster, Nucl.\ Phys.\ {\bf B232} (1984), 445 \\
J. Koller and P. van Baal, Nucl.\ Phys.\ {\bf B273} (1986), 387; Phys.\ Rev.\
Lett.\ {\bf 58} (1987), 2511; Nucl.\ Phys.\ {\bf B302} (1988), 1 \\
P. van Baal and J. Koller, Ann.\ Phys.\ (N.Y.) {\bf 174} (1987), 299 \\
P. van Baal, Nucl.\ Phys.\ {\bf B351} (1991), 183
\bibitem{lattice1} G. S. Bali, K. Schilling, A. Hulsebos, A. C. Irving,
C. Michael and P. W. Stephenson, Phys.\ Lett.\ {\bf B309} (1993), 378
\bibitem{peardon} M. Peardon, Nucl.\ Phys.\ {\bf B} (Proc.\ Suppl.) {\bf 63}
(1998), 22 \\
C. Morningstar, private communication to A.W.
\bibitem{dzublik} A. Ya.\ Dzublik, V. I. Ovcharenko, A. I. Steshenko and
G. F. Filippov, Yad.\ Fiz.\ {\bf 15} (1972), 487 [Sov.\ J. Nucl.\ Phys.\
{\bf 15} (1972), 487] \\
W. Zickendraht, J. Math. Phys. {\bf 10} (1969), 30; {\em ibid.}\ {\bf 12}
(1971), 1663
\bibitem{complset} O. Casta\~nos, A. Frank, E. Chac\'on, P. O. Hess and
M. Moshinsky, J. Math.\ Phys.\ {\bf 23} (1982), 2537
\bibitem{newvar} P. O. Hess, Phys.\ Rev.\ {\bf D40} (1989), 918 \\
P. O. Hess, J. C. L\'opez and C. R. Stephens, in ``Proceedings of the
Structure of Vacuum and Elementary Matter'', George, South Africa,
March 1996, ed.\ by H. St\"ocker, World Scientific (1997), p.\ 640
\bibitem{hamermesh} M. Hamermesh, ``Group Theory and its Application
to Physical Problems'', Diver Publications, New York (1989)
\bibitem{viollier} P. O. Hess and R. D. Viollier, Phys.\ Rev.\ {\bf D34}
(1986), 258
\bibitem{elliott} J. P. Elliott, Proc.\ Roy.\ Soc.\ {\bf A245} (1958), 128 and
562
\bibitem{wyborne} B. G. Wybourne, ``Symmetry Principles and Atomic
Spectroscopy'', Wiley, New York (1969)
\bibitem{wyborne2} B. G. Wybourne, computer program ``Schur'', version
1.2-IBM-PC (1984)
\bibitem{ramon} R. L\'opez, P. O. Hess, P. Rochford and J. P. Draayer,
J. Phys.\ {\bf G23} (1990), L229
\bibitem{edmonds} A. R. Edmonds, ``Angular Momentum in Quantum Mechanics'',
Princeton University Press, New Jersey (1974)
\bibitem{charge} R. Slansky, Phys.\ Rep.\ {\bf 79} (1981), 1 (see p.\ 43)
  and references therein \\
  A. Weber, PhD thesis, University of Heidelberg 1995 (unpublished)
\bibitem{eps} T. Molien, Berliner Sitzungsberichte {\bf 11} (1898), 1152 \\
R. Gaskell, A. Peccia and R. T. Sharp, J. Math.\ Phys.\ {\bf 19} (1978), 727
\bibitem{sharp} E. Chac\'on, M. Moshinsky and R. T. Sharp,
J. Math.\ Phys.\ {\bf 17} (1976), 668 \\
E. Chac\'on and M. Moshinsky, J. Math.\ Phys.\ {\bf 18} (1977), 870
\bibitem{epdsmir} Yu.\ I. Smirnov and P. O. Hess, work in progress
\bibitem{greiner} J. M. Eisenberg and W. Greiner, ``Nuclear Theory:
Nuclear Models'', 3rd edition, North-Holland, Amsterdam (1987)
\bibitem{kinam} E. Chac\'on, P. O. Hess and C. R. Sarma,
Kinam {\bf 4} (1982), 227
\bibitem{gilmore} R. Gilmore, ``Lie Groups, Lie Algebras and some
of their Applications'', John Wiley, New York (1974)
\bibitem{dragt} A. J. Dragt, J. Math.\ Phys.\ {\bf 6} (1965), 533
\bibitem{michael}
C. Michael, in ``Confinement, Duality, and Nonperturbative Aspects of QCD'',
ed.\ by P. van Baal, NATO ASI Series B 368, Plenum Press, New York (1998);
in ``Hadron Spectroscopy 1997'', ed.\ by S.-U. Chung and H. J. Willutzki,
AIP Conference Proceedings 432, American Institute of Physics (1998)
\bibitem{luo} X.-Q. Luo, Q. Chen, S. Guo, X. Fang and J. Liu, Nucl.\ Phys.\
{\bf B} (Proc.\ Suppl.) {\bf 53} (1997), 243
\bibitem{morning} C. Morningstar and M. Peardon, Phys.\ Rev.\ {\bf D56}
(1997), 4043
\bibitem{cern} CERN-routine Minuit, Cern library (1995)
\bibitem{bergb} B. Berg and A. Billoire, Nucl.\ Phys.\ {\bf B221} (1983), 109
\bibitem{operatordim} R. L. Jaffe, K. Johnson and Z. Ryzak, Ann.\ Phys.\
(N.Y.) {\bf 168} (1986), 344
\bibitem{sumrules} M. A. Shifman, A. I. Vainshtein and V. I. Zakharov, Nucl.\
Phys.\ {\bf B147} (1979), 385 and 448 \\
S. Narison, Nucl.\ Phys.\ {\bf B509} (1998), 312; Nucl.\ Phys.\ {\bf B}
(Proc.\ Suppl.) {\bf 64} (1998), 210
\bibitem{dittner} P. Dittner, Commun.\ Math.\ Phys.\ {\bf 22} (1971), 238;
{\em ibid.}\ {\bf 27} (1972), 44
\end{thebibliography}
\end{document}